\magnification=\magstephalf
\input amstex
\loadbold
\documentstyle{amsppt}
\refstyle{A}
\NoBlackBoxes

\vsize=7.5in

\def\pf{\hfill $\square$}
\def\c{\cite}

\def\fg{\frak{g}}
\def\fh{\frak{h}}

\def\end{\text{End}}

\def\bl{\boldsymbol \lambda}
\def\bn{\boldsymbol \nu}
\def\bL{\bold L}
\def\bM{\bold M}
\def\bbl{\boldkey l}

\def\bA{{\bold A}{\bold d}}

\topmatter
\title A family of hyperbolic spin Calogero-Moser systems and 
       the spin Toda lattices  \endtitle
\leftheadtext{L.-C. Li}
\rightheadtext{hyperbolic spin CM  and spin Toda}

\author Luen-Chau Li\endauthor
\address{L.-C. Li, Department of Mathematics,Pennsylvania State University
University Park, PA  16802, USA}\endaddress
\email luenli\@math.psu.edu\endemail 
\abstract  In this paper, we continue to develop a general scheme to
study a broad class of integrable systems naturally associated with the
coboundary dynamical Lie algebroids.  In particular, we present
a factorization method for solving the Hamiltonian flows.   We also 
present two important class of new
examples, a family of hyperbolic spin Calogero-Moser systems and
the spin Toda lattices.  To illustrate our factorization theory,
we show how to solve these Hamiltonian systems explicitly.

\endabstract
\endtopmatter

\document
\subhead
1. \ Introduction
\endsubhead

\baselineskip 15pt
\bigskip

In the theory of integrable systems, a wide range of important examples
are covered by the Adler-Kostant-Symes scheme and its generalization known as 
classical r-matrix theory  (see \c{A}, \c{K}, \c{S}, \c{RSTS1}, \c{RSTS2}, 
\c{AvM}, \c{STS1},\c{STS2}, \c{RSTS3}, \c{FT}, \c{LP1} and the references 
therein).  As is well-known, classical r-matrices are naturally
associated with Poisson structures on Lie groups and duals of Lie
algebras and the corresponding geometric objects have been used with great 
success in the solutions of many integrable Hamiltonian systems.

In the early 90's, dynamical analog of the classical r-matrices was
discovered in the study of Wess-Zumino-Witten (WZN) 
conformal field theory \c{BDF}, \c{F}.  Since then, these objects have 
cropped up in other areas as well (see, for example, \c{BAB}, \c{ABB},
\c{Lu}, \c{AM}) and their geometric
meaning was unraveled by Etingof and Varchenko in their fundamental
paper \c{EV}.  While
classical r-matrices play a role in Poisson Lie group theory \c{D}, the
authors in \c{EV} showed that an appropriate geometric setting for the
classical dynamical r-matrices is that of a special class of Poisson
groupoids (a notion due to Weinstein \c{W1}), the so-called coboundary 
dynamical Poisson groupoids.
If $R$ is an $H$-equivariant classical dynamical r-matrix, and 
$(\Gamma, \{\,\cdot,\cdot \,\}_{R})$ is the associated coboundary
dynamical Poisson groupoid, then it follows from Weinstein's 
coisotropic calculus \c{W1} or otherwise that the Lie algebroid
dual $A^{*}\Gamma$ also has a natural Lie algebroid structure
\c{LP2}, \c{BKS}.  We shall call $A^{*}\Gamma$ the coboundary dynamical
Lie algebroid associated to $R$ and it is this class of Lie algebroids
which we use in the study of integrable systems in \c{LX2} and in the
present work.

Our purpose in this paper is twofold.  First of all, we will continue to
develop a general scheme (which we initiated in \c{LX2}) to study
integrable systems based on realization in the dual bundles 
of coboundary dynamical Lie algebroids.  To summarize, the class of 
invariant Hamiltonian systems which admits such a realization 
(for the genuinely dynamical case) has the following key features: 
(a) the systems are 
defined on a Hamiltonian $H$-space
$X$ with equivariant momentum map $J$ and the Hamiltonians are the
pull-back of natural invariant functions by an $H$-equivariant 
realization map, (b) the pullback of natural invariant functions
do not Poisson commute everywhere on $X$, but they do so on a fiber
$J^{-1}(\mu)$ of the momentum map, (c) the reduced Hamiltonian
systems on $X_{\mu} = J^{-1}(\mu)/H_{\mu}$ ($H_{\mu}$ is the isotropy
subgroup at $\mu$) admit a natural collection of Poisson commuting
integrals.  In this work, our main focus is on the case in which $R$ 
is a solution of the modified dynamical Yang-Baxter equation 
(mDYBE).  As we pointed out in the announcement \c{L1}, the (mDYBE)
is associated with a factorization problem on the trivial Lie groupoid
$\Gamma$.  By making use of the algebraic and geometric structures
associated with (mDYBE) (which will be fully worked out here),
we will develop an effective  method to
integrate the Hamiltonian flows on $J^{-1}(\mu)$ (which parallels
the one announced in \c{L1} for the groupoid framework) based on
this factorization.   Hence we can
obtain the integrable flows on $X_{\mu}$ by reduction.

Our second purpose in this work is to give two important class of
new examples and to illustrate our factorization theory using these
examples.  Our first class of examples is a family of hyperbolic
spin Calogero-Moser (CM) systems and their associated integrable models,
corresponding to the solutions of (mDYBE) for pairs $(\fg,\fh)$ of
Lie algebras, as classified in \c{EV}.  Here, $\fg$ is simple, and
$\fh \subset \fg$ is a Cartan subalgebra.  As such, our systems are
parametrized by subsets $\pi^{\prime}$ of a simple system of roots
$\pi$.  Note that in the special case where $\pi^{\prime} = \pi$, our 
corresponding integrable model is isomorphic to the one in \c{R}, and the 
$sl(N)$ case has also appeared in \c{AB}, \c{KBBT},for example 
(see Remark 5.5).
The second class of examples was actually discovered when an appropriate 
scaling limit is applied to the hyperbolic spin CM systems (the ones which 
are not integrable).  Remarkably, the obstruction to integrability
dissolves in the scaling limit, leading to a family of integrable models
which we will call the spin Toda lattices (again, these are parametrized
by subsets $\pi^{\prime}$ of a simple system).  As it turns out, the spin 
Toda lattices are systems which admit realization in the dual bundle 
${\bold A}$
of the coboundary dynamical Lie algebroid ${\bold A}^{*} \simeq
T\fh \times \fg$ associated to the standard r-matrix and the reduction
of these $H$-invariant systems lead to a family of Toda lattices parametrized
by $\pi^{\prime}$.
So this gives us a first nontrivial example in which a constant r-matrix 
is relevant.  

The paper is organized as follows.  In Section 2, we derive an intrinsic
expression for the Lie-Poisson structure on the dual bundle of a coboundary
dynamical Lie algebroid which is important for subsequent developments of
our program.  We also give the complete set of equations for a natural
class of invariant Hamiltonian systems.  In Section 3, we reprove 
(essentially) Theorem 3.10  in \c{LX2} using an intrinsic point of view, 
without having to assume the existence of an $H$-equivariant map 
$g:X\longrightarrow H$ (also we do not assume $\fh = Lie (H)$ is Abelian).
We also compute how the realization map
evolves under our invariant Hamiltonian systems on $X$ based on the 
development in Section 2.  As the reader will see, the integrable
flows on the reduced space $X_{\mu}$ are actually realized on a
Poisson quotient of a coisotropic submanifold of $A\Gamma$, which
in some sense is the analog of the gauge group bundle in \c{L1}.
In Section 4, we discuss the algebraic and geometric structures
associated with (mDYBE), leading up to a factorization method for
solving the Hamiltonian flows.  In Section 5, we introduce a family
of hyperbolic spin Calogero-Moser systems using Proposition 4.2 (a)
and consider the associated integrable models.  Then we consider 
scaling limits of the hyperbolic spin CM systems at the levels of
the Hamiltonians, the equations of motion and the (generalized) Lax 
equations.  At the end of the section, we work out the realization
picture for the spin Toda lattices and also consider their reduction.
Section 6 is concerned with the solution of the hyperbolic spin
CM systems and the spin Toda lattices, utilizing the factorization
method of Section 4.  Here, the reader will see how the concrete
factorization problems are being solved.  In a remark, we will also
discuss the solution of a family of hyperbolic spin Ruijenaars-Schneider 
models (introduced in \c{L1} and related to the affine Toda field 
theories \c{BH}) in the general case.  We shall address the 
complete integrability and other
aspects of the integrable models here in subsequent publications.
For the solution of the systems in \c{LX2} using the method developed
here, we refer the reader to the forthcoming work \c{L2} (see also
Remark 5.5 (c)).
\smallskip
{\bf Acknowledgments.} The author would like to thank the referee
for a helpful question which has led him to put back a missing keyword in the 
formulation of several results in Section 4.  Special thanks are also due 
to Reeva Goldsmith for converting the AMS-TeX file to LaTeX.

\bigskip
\bigskip

\subhead
2. \ Coboundary dynamical Lie algebroids and Lie-Poisson structures
\linebreak \phantom{faak}\,on their dual bundles
\endsubhead

\bigskip
In this section, our main goal is to derive an intrinsic formula for the 
Lie-Poisson structure on the dual bundle of a coboundary dynamical Lie
algebroid which is important for subsequent developments.  As the
reader will see, the same method of calculation is also used in
Section 5 to write down the Lie-Poisson structure associated with
a trivial Lie algebroid, whose vertex Lie algebra is given
by a semi-direct product.

We begin by recalling the definition of a Lie algebroid.

\definition
{Definition 2.1} A Lie algebroid over a manifold $M$ is a smooth vector bundle
$\pi_{A} : A\longrightarrow M$ equipped with a Lie bracket $[\cdot,\cdot]_{A}$
on its space $Sect (M, A)$ of smooth sections and a bundle map
$a_{A} : A\longrightarrow TM$ (called the anchor map) such that
\smallskip
\noindent (a) the bundle map $a_A$ induces a Lie algebra homomorphism 
$Sect (M,A)\longrightarrow Sect (TM)$ (which we also denote by $a_A$),
\smallskip
\noindent (b) for any $X$, $Y\in Sect (M,A)$ and $f\in C^{\infty}(M)$,
the Leibnitz identity 
$$[\,X, fY\,]_{A} = f[\,X,Y\,]_{A} +(a(X)f)Y$$
holds.
\enddefinition

Let $(A, [\cdot,\cdot]_{A},a_{A})$ be a Lie algebroid over $M$, and let
$\pi_{A^*}:A^{*}\longrightarrow M$ be its dual bundle.  For any 
$X\in Sect (M,A)$, we can associate a smooth function $l_{X}$
on $A^{*}$ by putting $l_{X} (\xi)=<\xi, X\circ \pi_{A^*}(\xi)>$
for all $\xi \in A^{*}$.

\proclaim
{Theorem 2.2 \c{CDW}}  There exists a unique Poisson structure on
$A^{*}$ (called the Lie-Poisson structure) which is characterized
by the property
$$\{\,l_{X}, l_{Y}\,\} = l_{[\,X,Y\,]_{A}}$$
for all $X$, $Y\in Sect (M,A)$.
\endproclaim

\smallskip
\noindent {\bf Remark 2.3} In \c{C} and \c{W2},there are two extra conditions 
in the 
characterization of the Lie-Poisson structure, namely, 
$\{f\circ \pi_{A^*}, g\circ \pi_{A^*}\} 
=0$, and $\{l_{X}, f\circ \pi_{A^*}\}= (a(X)f)\circ \pi_{A^*}$ for all 
$f$, $g\in
C^{\infty}(M)$, and $X\in Sect(M,A)$.  However, it is not hard to show that 
these are actually consequences of $\{\,l_{X}, l_{Y}\,\} = l_{[X,Y]_{A}}$ and 
the
properties of the Lie algebroid bracket $[\cdot,\cdot]_{A}$.  We shall
use these two conditions in an essential way in (2.10) below.

\smallskip
We now recall the definition of a coboundary dynamical Lie algebroid.
Let $G$ be a connected Lie group, and $H\subset G$ a connected Lie
subgroup.  We shall denote by $\fg$ and $\fh$ the corresponding Lie
algebras and let $\iota :\fh \longrightarrow \fg$ be the Lie inclusion.
Let $U\subset \fh^*$ be a connected 
$Ad_H^*$-invariant open subset, and let  
$R:U\longrightarrow L(\fg^*, \fg)$ be a classical dynamical r-matrix
(here and henceforth we denote by
$L(\fg^*,\fg)$ the set of linear maps from $\fg^*$ to $\fg$), i.e.
$R$ is pointwise  skew symmetric
$$<R(q)(A), B>=- <A, R (q) B>\eqno (2.1)$$
and satisfies the classical dynamical Yang-Baxter condition
$$\eqalign {&[R(q)A, R(q)B] +R(q)(ad^*_{R(q)A}B-ad^*_{R(q)B}A)\cr
+&dR(q)\iota^*A(B) - dR(q)\iota^*B(A) + d<R(A),B>(q) 
= \chi (A,B),\cr} \eqno (2.2)$$
for all $q\in U$, and all $A, B\in \fg^*$, 
where $\chi : \fg^*\times \fg^* \longrightarrow  \fg$ is
$G$-equivariant.
\smallskip
The dynamical $r$-matrix is said to be $H$-equivariant
if and only if
$$R(Ad^*_{h^{-1}} q)=Ad_h \circ R(q)\circ Ad^*_h \eqno (2.3)$$
for all $h\in H, q\in U.$  We shall equip 
$\Gamma=U\times G\times U$ with the trivial Lie groupoid
structure over $U$ \c{M} with target and source maps
$$\alpha (u,g,v)= u, \quad  \beta(u,g,v)= v \eqno(2.4)$$
and multiplication map
$$m((u,g,v), (v,g',w))= (u,gg',w).\eqno(2.5)$$

Recall that associated with an $H$-equivariant classical dynamical 
r-matrix $R$ is a 
natural
Poisson structure $\{\,\cdot, \cdot \,\}_R$ on $\Gamma$ \c{EV}
such that the pair $(\Gamma, \{\,\cdot, \cdot \,\}_R)$ is a Poisson groupoid
in the sense of Weinstein \c{W1} (see \c{L1} for the intrinsic forms
of $\{\,\cdot, \cdot \,\}_R$).  Let $A \Gamma := \bigcup_{q \in U} T_
{\epsilon(q)} \alpha ^{-1} (q) = \bigcup_{q\in U} \lbrace 0_{q} \rbrace
\times \fg \times \fh^*$ be the Lie algebroid of $\Gamma.$  Then by Weinstein's
coisotropic calculus \c{W1} or otherwise, the Lie algebroid dual $A^*\Gamma
= \bigcup_{q\in U} \lbrace 0_{q} \rbrace \times \fg^* \times \fh$ also
has a natural Lie algebroid structure \c{BKS},\c{LP2}  such that the pair
$(A\Gamma, A^*\Gamma)$ is a Lie bialgebroid in the sense of Mackenzie
and Xu \c{MX}.  We shall denote the Lie brackets on $Sect(U,A\Gamma)$
and $Sect (U,A^*\Gamma)$ respectively by $[\cdot, \cdot]_{A\Gamma}$
and $[\cdot, \cdot]_{A^*\Gamma}.$  Throughout the paper, the pair
 $(A^{*}\Gamma, [\cdot,\cdot]_{A^{*}})$
together with the anchor map $a_{*}:A^{*}\Gamma\longrightarrow TU$
given by
$$a_{*} (0_q,A,Z)=(q, \iota^{*}A-ad^{*}_{Z}q)\eqno(2.6)$$
will be called the coboundary dynamical Lie algebroid associated to $R$.
Explicitly, the Lie bracket $[\cdot, \cdot]_{A^*\Gamma}$ on 
$Sect(U,A^{*}\Gamma)$ is given by the following expression \c{BKS},\c{LP2}:
$$\eqalign{& [(0,A,Z),(0,A',Z')]_{A^{*}\Gamma} (q)\cr
     = & (0_q, dA'(q)(\iota^{*}A(q)-ad^{*}_{Z(q)} q)
         -dA(q)(\iota^{*} A'(q)-ad^{*}_{Z'(q)} q)\cr
       &\quad -ad^{*}_{R(q)A(q)-Z(q)} A'(q)
         +ad^{*}_{R(q)A'(q)-Z'(q)} A(q),\cr
       &\quad \,dZ'(q)(\iota^{*}A(q)-ad^{*}_{Z(q)} q)
         -dZ(q)(\iota^{*} A'(q)-ad^{*}_{Z'(q)} q)\cr
       &\quad -[Z,Z'](q) + <dR(q)(\cdot)A(q),A'(q)>)\cr}\eqno(2.7)
$$
where $A,A':U\longrightarrow \fg^{*}$, $Z,Z':U\longrightarrow \fh$ are
smooth maps and $<dR(q)(\cdot)A(q),A'(q)>$ is the element in $\fh$
whose pairing with $\lambda \in \fh^*$ is $<dR(q)(\lambda)A(q),A'(q)>.$ 

In the rest of the section, we shall make the identifications
$$A\Gamma\simeq U\times \fh^*\times \fg, \quad
  A^{*}\Gamma\simeq U\times \fh\times \fg^* .\eqno(2.8)
$$

Let us fix a point $(q,\lambda,X)\in A\Gamma$.  In order to derive an
intrinsic expression for the Lie-Poisson bracket
$\{\varphi,\psi\}_{A\Gamma} (q,\lambda,X)$  on the dual bundle $A\Gamma$
of the coboundary dynamical Lie algebroid 
$(A^{*}\Gamma, [\cdot,\cdot]_{A^{*}\Gamma},a_{*})$, we need to 
introduce some notation. 
To start with, let $Pr_{i}$ be the 
projection map onto the $i$-th factor of 
$U\times \fh^{*}\times \fg \simeq A\Gamma$,
$i=1,2,3.$
If $\varphi \in C^{\infty} (A\Gamma)$, we have 
$d \varphi (q,\lambda,X) = (\delta_{1}\varphi,\delta_{2}\varphi,
\delta \varphi)$, where the partial derivatives are defined by
$$\eqalign {&<\delta_{1} \varphi, \mu>={d\over dt}_{|_{t=0}}
\varphi (q+t\mu, \lambda,X),\quad <\delta_{2} \varphi, \mu>
={d\over dt}_{|_{t=0}} \varphi (q,\lambda+t\mu,X),\,
\mu\in \fh^*\cr
& <\delta \varphi, Y>={d\over dt}_{|_{t=0}} \varphi (q,\lambda, X+tY),\,
Y\in \fg.\cr}$$
We also associate with $\varphi$ the function $\widetilde \varphi$ on
$U$, defined by $\widetilde \varphi (u) = \varphi (u,\lambda,X)$.
On the other hand, $s(\varphi): U\longrightarrow U\times \fh\times \fg^{*}$
will denote the constant section of $U\times \fh\times \fg^{*}$ given
by $s(\varphi) (u) =(u, \delta_{2}\varphi, \delta \varphi)$, where
$\delta_{2}\varphi$, $\delta \varphi$ are the partial derivatives
evaluated  at the fixed point $(q,\lambda,X)$.  

Now, it is easy to check by a direct calculation that 
$d({\widetilde \varphi}\circ Pr_{1}) (q,\lambda,X)=(\delta_{1}\varphi,
0,0)$, while 
$d l_{s(\varphi)} (q,\lambda,X)=(0,\delta_{2}\varphi,\delta\varphi)$.
Thus we have
$$d \varphi(q,\lambda,X)=d(l_{s(\varphi)} +{\widetilde \varphi}\circ
Pr_{1}) (q, \lambda,X).\eqno(2.9)$$
Therefore,
$$\eqalign {& \{\varphi, \psi\}_{A\Gamma} (q,\lambda,X)\cr
          = & \{l_{s(\varphi)}+{\widetilde \varphi}\circ Pr_{1},
                l_{s(\psi)}+{\widetilde \psi}\circ Pr_{1} \}_{A\Gamma}
                (q,\lambda,X)\cr
          = & l_{[s(\varphi),s(\psi)]_{A^{*}\Gamma}} (q,\lambda,X)
              +d{\widetilde \psi}(q) a_{*}(s(\varphi)) (q)\cr
            &-d{\widetilde \varphi}(q) a_{*}(s(\psi))(q).\cr}\eqno(2.10)$$
By using the expression for  $[\cdot,\cdot]_{A^{*}\Gamma}$ in (2.7), we
have
$$
\aligned
       & l_{[s(\varphi),s(\psi)]_{A^{*}\Gamma}} (q,\lambda,X) \\
       = &<\lambda, -[\delta_{2}\varphi, \delta_{2}\psi]
          + <dR(q)(\cdot)\delta \varphi, \delta \psi>>\\
         & +<X, -ad^{*}_{R(q)\delta \varphi-\delta_{2}\varphi}
          \delta \psi + ad^{*}_{R(q)\delta \psi-\delta_{2}\psi}
          \delta \varphi >.
\endaligned
$$
Meanwhile, from the expression for the anchor map $a_{*}$, we find
$$ d{\widetilde \varphi}(q) a_{*}(s(\varphi))(q)
  =<\delta_{1} \varphi, \iota^{*}\delta \psi- ad^{*}_{\delta_{2}\psi}q>.$$
Assembling the calculations, we have the following result.

\proclaim
{Theorem 2.4} The Lie-Poisson structure on the dual bundle $A\Gamma$
of the coboundary dynamical Lie algebroid 
$(A^{*}\Gamma, [\cdot,\cdot]_{A^{*}\Gamma},a_{*})$
is given by
$$\eqalign{& \{\varphi, \psi\}_{A\Gamma} (q,\lambda,X)\cr
  =&-<\lambda, [\delta_{2}\varphi,\delta_{2}\psi]>+
     <dR(q)(\lambda)\delta \varphi,\delta\psi>\cr
     & +<X, -ad^{*}_{R(q)\delta \varphi-\delta_{2}\varphi}
          \delta \psi + ad^{*}_{R(q)\delta \psi-\delta_{2}\psi}
          \delta \varphi >\cr
     &-<q,[\delta_{2}\varphi,\delta_{1}\psi]+[\delta_{1}\varphi,
       \delta_{2}\psi]>+<\delta_{1}\psi, \iota^{*}\delta \varphi>
       -<\delta_{1}\varphi,\iota^{*}\delta\psi>.\cr}$$
\endproclaim

\noindent{\bf Remark 2.5}  In a similar fashion, we can show that the
Lie-Poisson bracket on the dual bundle $A^{*}\Gamma$ of the trivial
Lie algebroid $(A\Gamma, [\cdot,\cdot]_{A\Gamma},a)$ is given by
$\{\varphi, \psi\}_{A^{*}\Gamma}(q,p,\xi)=<\delta_{2}\varphi,\delta_{1}\psi>
-<\delta_{1}\varphi, \delta_{2}\psi>+ <\xi,[\delta\varphi,\delta\psi]>.$
The reader is referred to  Proposition 5.10 below for the details of
a similar calculation.

If $(P, \{\cdot,\cdot \}_{P})$ is a Poisson manifold, then for each
$f\in C^{\infty}(P)$, we shall define the associated Hamiltonian
vector field $X_{f}$ using the convention 
$X_{f}.g = \{\,f \, ,g\,\}_{P}$.

\proclaim
{Corollary 2.6} The Hamiltonian vector field on $A\Gamma$ associated
to $\varphi \in C^{\infty} (A\Gamma)$ is given by
$$
\aligned
       & X_{\varphi} (q,\lambda,X)\\
     = &(\iota^{*}\delta \varphi - ad^{*}_{\delta_{2}\varphi} q,
         -ad^{*}_{\delta_{2} \varphi} \lambda
         +\iota^{*}ad^{*}_{X} \delta \varphi
         -ad^{*}_{\delta_{1}\varphi} q,\\
       & \quad
         [X, R(q)\delta \varphi -\delta_{2}\varphi]
         + dR(q)(\lambda) \delta \varphi - \delta_{1} \varphi
         +R(q)(ad^{*}_{X} \delta \varphi)).
\endaligned
$$
\endproclaim

Now, a  natural collection of invariant functions on $A\Gamma$ is
$Pr^{*}_{3} I(\fg)$, where $I(\fg)$ is the ring of ad-invariant
functions on $\fg$.  The following result is an easy consequence
of Theorem 2.4  and Corollary 2.6.

\proclaim
{Corollary 2.7} (a) The Hamilton's equation on $A\Gamma$ generated by
$Pr^{*}_{3} f$, $f\in I(\fg)$ is of the form
$$
\aligned
       &\dot q = \iota^{*} df(X),\\
       &\dot \lambda = 0,\\
       &\dot X = [X, R(q) df(X)] +  dR(q)(\lambda)(df(X)).
\endaligned
$$
\smallskip
\noindent (b) For all $f_1$, $f_2\in I(\fg)$, we have
$$
\aligned
       & \{\,Pr^{*}_{3} f_{1}, Pr^{*}_{3} f_{2}\,\}_{A\Gamma} (q,\lambda,X)\\
     = & < dR(q)(\lambda)(df_{1}(X)), df_{2}(X)>.
\endaligned
$$
\endproclaim

\noindent{\bf Remark 2.8}  If $R$ is a constant r-matrix, then it is
immediate from Corollary 2.7 (b) above that functions in 
$Pr_{3}^{*} I(\fg)$ Poisson commute on $A\Gamma$.  In this case,
the equation for $X$ in part (a) of the same corollary  is a Lax equation 
in the standard r-matrix framework for Lie algebras \c{STS1}.
So when $R$ is constant, what we have here is a slight extension of the
standard framework. For an example associated with a constant r-matrix
which fits into our framework, we refer the reader to Section 5.

\bigskip
\bigskip

\subhead
3. \ Realization of Hamiltonian systems in coboundary dynamical Lie 
\linebreak \phantom{faak}\, algebroids
\endsubhead

\bigskip

Let $(A^{*}\Gamma, [\cdot,\cdot]_{A^{*}\Gamma}, a_{*})$ be the
coboundary dynamical Lie algebroid corresponding to $R$, and let
$\rho: X\longrightarrow A\Gamma$ be a realization of a Poisson
manifold $(X, \{\,\cdot,\cdot\,\}_{X})$ in the dual bundle 
$A\Gamma$ of the Lie algebroid $A^{*}\Gamma$, i.e., $\rho$ is
a Poisson map. If $Pr_{i}$ is the projection map onto the $i$-th factor of 
$U\times \fh^{*}\times \fg \simeq A\Gamma$,
$i=1,2,3,$ we put

$$m = Pr_{1}\circ \rho: X \longrightarrow U, \eqno(3.1)$$
$$\tau = Pr_{2}\circ \rho: X \longrightarrow \fh^{*},\eqno(3.2)$$
$$L = Pr_{3}\circ \rho: X \longrightarrow \fg.\eqno(3.3)$$

We shall make the following assumptions:
\medskip

\noindent A1. $X$ is a Hamiltonian $H$-space with an equivariant
momentum map $J:X\longrightarrow \fh^*$,\newline
\noindent A2. the realization map $\rho$ is $H$-equivariant, where
$H$ acts on $A\Gamma$ via the formula

$$h\cdot(q,\lambda,X)=(Ad^{*}_{h^{-1}}q, Ad^{*}_{h^{-1}}\lambda,Ad_{h}X),
\eqno(3.4)
$$

\noindent A3. for some regular value $\mu\in \fh^{*}$ of $J$, 

$$\rho (J^{-1} (\mu)) \subset U\times \{0\}\times \fg. \eqno(3.5)$$

Note that the condition in (3.4) is the natural generalization of
the corresponding condition in \c{LX2} since we do not assume $\fh$
is Abelian here.  On the other hand, our assumption A3 is stronger
than what we had in \c{LX2}.  Our purpose in this section is to
exhibit the intrinsic role played by the orbit space
$(U\times \{0\}\times \fg)/H$ of the action in (3.4) in the
reduction to integrable flows.  We also compute how the
realization map evolves under our invariant Hamiltonian systems
on $X$.

\proclaim
{Proposition 3.1}  With the action defined in (3.4), the
dual bundle $A\Gamma$ of the coboundary dynamical Lie algebroid
$A^{*}\Gamma$ equipped with the Lie-Poisson structure is a Hamiltonian 
$H$-space with equivariant momentum
map $\gamma:A\Gamma\longrightarrow \fh^{*}$, $(q,\lambda,X)\mapsto
\lambda.$
\endproclaim

\demo
{Proof}  Denote the action by $\Phi$.  If $\varphi\in C^{\infty}(A\Gamma)$,
it follows by a direct calculation that
$\delta_{i}(\varphi\circ\Phi_{h})(q,\lambda,X)
= Ad_{h^{-1}} \delta_{i}\varphi(\Phi_{h}(q,\lambda,X))$, $i=1,2$
and $\delta(\varphi\circ\Phi_{h})(q,\lambda,X)
= Ad^{*}_{h}\delta\varphi(\Phi_{h}(q,\lambda,X)).$  The
assertion that $\Phi_{h}$ is Poisson then follows upon using
the formula in Theorem 2.4 and the fact that $R$ is $H$-equivariant.
Now, for any $Z\in\fh$, we have
$${d\over dt}_{|_{t=0}} \Phi_{e^{tZ}} (q,\lambda,X)= 
(-ad^{*}_{Z}q,-ad^{*}_{Z}\lambda,
ad_{Z}X).$$
Comparing the right hand side of the above expression with the formula
in Corollary 2.6, it is clear that this is equal to 
$X_{\widehat \gamma (Z)} (q,\lambda,X)$, where
${\widehat \gamma (Z)} (q,\lambda,X)= <\lambda,Z>$.  Hence
$\gamma (q,\lambda,X)= \lambda.$
\pf
\enddemo

From this result, it follows that $X=A\Gamma$, and $\rho=id_{A\Gamma}$
satisfy assumptions A1-A3 above with $\mu=0$ and we have 
$\gamma^{-1}(0)= U\times \{0\}\times\fg.$

We shall denote by $H_{\mu}$ the isotropy subgroup of $\mu$ for the
$H$-action on $X$.  Then it follows by Poisson reduction \c{MR},
\c{OR} (see \c{OR} for the singular case) that the variety
$X_{\mu} = J^{-1}(\mu)/ H_{\mu}$ inherits a unique Poisson structure 
$\{\,\cdot,\cdot \,\}_{X_{\mu}}$
satisfying 

$$\pi_{\mu}^* \{f_1, f_2 \}_{X_{\mu}} = i_{\mu}^* \{\widetilde{f_1},
\widetilde{f_2} \}_X .\eqno(3.6)$$
Here, $i_{\mu}:J^{-1} (\mu) \longrightarrow X$ is the inclusion map,
$\pi_{\mu}:J^{-1} (\mu)\longrightarrow X_{\mu}$ is the canonical projection,
$f_1$, $f_2 \in C^{\infty} (X_{\mu})$, and $\widetilde{f_1}$,
$\widetilde{f_2}$ are (locally defined) smooth extensions of 
$\pi_{\mu}^* f_1$, $\pi_{\mu}^* f_2$ with differentials vanishing
on the tangent spaces of the $H$-orbits.  For the case where
$X=A\Gamma$, $\rho=id_{A\Gamma}$, it is clear that the isotropy subgroup
at $\mu=0$ is $H$ itself and so we have the Poisson variety

$$(A\Gamma_0 = {\gamma^{-1}(0)/ H}, \{\,\cdot,\cdot \,\}_{A\Gamma_0}),
\eqno(3.7)$$
with the inclusion map $i_H  :\gamma^{-1}(0) \longrightarrow A\Gamma$ and
the canonical projection $\pi_H :\gamma^{-1}(0) \longrightarrow A\Gamma_0.$

Clearly, functions in $i_H^* Pr_3^* I(\fg) \subset C^{\infty} (\gamma^{-1}(0))$
are $H$-invariant, hence they descend to functions in 
$C^{\infty} (A\Gamma_0)$.  On the other hand, it follows from
assumption A2 that the functions in $i_{\mu}^* L^* I(\fg) \subset
C^{\infty} (J^{-1}(\mu))$ drop down to functions in $C^{\infty} (X_{\mu}).$  
Now,
by assumption A2-A3, and the fact that $\rho$ is Poisson, it follows
from \c{OR} that $\rho$ induces a unique Poisson map

$$\widehat{\rho}: X_{\mu} \longrightarrow A\Gamma_0 = {(U\times \{0\}\times\fg)
/ H} \eqno(3.8)$$
characterized by  $\pi_H \circ \rho \circ i_{\mu}=\widehat{\rho}
\circ \pi_{\mu}.$  Hence $X_{\mu}$ admits a realization in the
Poisson variety $A\Gamma_0$.

We shall use the following notation.  For $f \in I(\fg)$, the unique
function in $C^{\infty} (A\Gamma_0)$ determined by $i_H^* Pr_3^* f$
will be denoted by $\bar f$; while the unique function in
$C^{\infty} (X_{\mu})$ determined by $i_{\mu}^* L^* f$ will be
denoted by ${\Cal F}_{\mu}$.  From the definitions, we have

$${\Cal F}_{\mu} \circ \pi_{\mu} = ({\widehat{\rho}}^{*} \bar f)
  \circ \pi_{\mu} = i_{\mu}^{*}L^{*} f\eqno(3.9)$$

\proclaim
{Theorem 3.2} Let $(X, \{\,\cdot,\cdot \,\}_X )$ be a Poisson manifold which
admits a realization $\rho: X \longrightarrow A\Gamma$ and assume
A1-A3 are satisfied.  Then there exist a unique Poisson structure
$\{\,\cdot,\cdot \,\}_{X_{\mu}}$ on the reduced space $X_{\mu} = 
J^{-1}(\mu)/H_{\mu}$
and a unique Poisson map $\widehat{\rho}$ such that

\noindent (a) for all $f_1$, $f_{2}\in I(\fg)$, $x\in J^{-1} (\mu)$,
we have

$$
\aligned
       & \{{\widehat{\rho}}^{*}{\bar f}_1, {\widehat{\rho}}^{*}
         {\bar f}_2 \}_{X_{\mu}}\circ \pi_{\mu} (x) \\
      =& <L(x), -ad^{*}_{R(m(x))df_{1}(L(x))} df_{2}(L(x))
          +ad^{*}_{R(m(x))df_{2}(L(x))} df_{1}(L(x)>.
\endaligned
$$
\smallskip
\noindent (b) functions ${\widehat{\rho}}^{*}{\bar f}$, $f\in I(\fg)$, 
Poisson commute in $(X_{\mu}, \{\,\cdot,\cdot \,\}_{X_{\mu}})$,
\smallskip
\noindent (c) if $\psi_t$ is the induced flow on 
$\gamma^{-1}(0)=U\times\{0\}\times \fg$ 
generated by the Hamiltonian $Pr^*_{3} f$, $f \in I(\fg)$, and 
$\phi_t$ is the Hamiltonian flow of ${\Cal F}=L^{*} f$ on $X$, then
under the flow $\phi_{t}$, we have
$$
\aligned
       & {d\over dt} m(\phi_{t}) = \iota^{*}
         df(L(\phi_{t})),\\
       & {d\over dt} \tau(\phi_{t}) = 0, \\
       & {d\over dt} L(\phi_{t}) = [\,L(\phi_{t}), R(m(\phi_{t}))
          df(L(\phi_{t}))\,] + dR(m(\phi_{t}))(\tau(\phi_{t}))
          df(L(\phi_{t}))
\endaligned
$$
where the term involving $dR$ drops out on $J^{-1}(\mu)$.
Moreover, the  reduction $\phi^{red}_t$ of $\phi_{t} \circ i_{\mu}$ on
$X_{\mu}$ defined by $\phi^{red}_t \circ \pi_{\mu} = \pi_{\mu} \circ
\phi_{t} \circ i_{\mu}$ is a Hamiltonian flow of ${\Cal F}_{\mu} =
{\widehat {\rho}}^{*} {\bar f}$ and ${\widehat {\rho}}\circ \phi^{red}_{t}
(\pi_{\mu} (x)) = \pi_{H} \circ \psi_{t} (\rho(x))$, \quad $x \in J^{-1}
(\mu)$.
\endproclaim

\demo
{Proof} (a) Since $\rho (J^{-1} (\mu)) \subset U\times \{0\}\times \fg$,
we have $\tau(x)=0$ for $x\in J^{-1}(\mu).$  Therefore,
$$
\aligned
       &\{{\widehat{\rho}}^{*}{\bar f}_1, {\widehat{\rho}}^{*}
         {\bar f}_2 \}_{X_{\mu}}\circ \pi_{\mu} (x) \\
     =\,& \{{\bar f}_1, {\overline f}_{2} \}_{A\Gamma_0} \circ \pi_{H}
         (\rho (x)) \\
     =\,&\{Pr^{*}_{3} f_1, Pr^{*}_{3} f_2 \}_{A\Gamma}
         (\rho (x)) \\
     =\,&<L(x), -ad^{*}_{R(m(x))df_{1}(L(x))} df_{2}(L(x))
          +ad^{*}_{R(m(x))df_{2}(L(x))} df_{1}(L(x)>
\endaligned
$$
where in the last step we have invoked the formula in Theorem 2.4
and the vanishing of $\tau(x)$ for $x\in J^{-1}(\mu)$.
\newline
(b) This is clear from part (a).
\newline
(c) Since $\rho$ is a Poisson map, we have 
${d\over dt} \rho(\phi_{t}) = X_{f\circ Pr_{3}}
(\rho(\phi_{t}))$ from which the equations follow
on invoking Corollary 2.7.  Finally, the assertion on $\phi^{red}_t$ is
basically a corollary of Theorem 2.16 of \c{OR} and the
relation  $\rho \circ \phi_{t} \circ i_{\mu} = \psi_{t} \circ \rho \circ
i_{\mu}$.
\pf
\enddemo

\noindent{\bf Remark 3.3}  In \c{LX2}, we have only written down the
equation for $L$ under the Hamiltonian flow $\phi_{t}$ (in the Abelian
case).  However, the full set of equations is important.  See Section 4
and Section 6 below.

\bigskip
\bigskip

\subhead
4. \ Factorization problems on Lie groupoids and exact solvability
\endsubhead

\bigskip
We shall develop a factorization method to solve the
(generalized) Lax equations in Corollary 2.7 (a) on the level set 
$\gamma^{-1}(0)$ of the momentum map $\gamma$.  For the first
part of this section, we shall use 
$A \Gamma  = \bigcup_{q\in U} \lbrace 0_{q} \rbrace
\times \fg \times \fh^*$, 
$A^*\Gamma
= \bigcup_{q\in U} \lbrace 0_{q} \rbrace \times \fg^* \times \fh$ ,
and when $\fg$ has an ad-invariant non-degenerate pairing, we shall
identify the Lie algebras with their duals.

As in \c{L1}, we introduce the bundle map
$${\Cal R}: A^*\Gamma \longrightarrow A\Gamma, 
(0_{q}, A, Z)\mapsto (0_{q}, -\iota Z+R(q)A, \iota^*A-ad^*_{Z}q)
\eqno(4.1)$$
and call it the {\sl r-matrix of the Lie algebroid} $A^{*}\Gamma$.
Also, we assume $R$ satisfies the modified dynamical Yang-Baxter
equation (mDYBE):
$$\eqalign {& [R(q)A, R(q)B]+R(q)(ad^*_{R(q)A}B-ad^*_{R(q)B}A)\cr
+\,&dR(q)\iota^*A(B) - dR(q)\iota^*B(A) + d<R(A),B>(q)\cr
=\,& - [K(A), K(B)]\cr}\eqno(4.2)$$
where $K\in L(\fg^*,\fg)$ is a nonzero symmetric map which
satisfies $ad_X\circ K + K\circ ad^*_X = 0$ for all $X \in \fg$,i.e,
$K$ is $G$-equivariant.

The next two results were announced in \c{L1}.  We give details of
the proof here.

\proclaim
{Lemma 4.1} If R satisfies (mDYBE), then the r-matrix
${\Cal R}: A^*\Gamma \longrightarrow A\Gamma$ satisfies the equation

$$\eqalign {& [{\Cal R}(0,A,Z),{\Cal R}(0,A',Z')]_{A\Gamma}-
{\Cal R}[(0,A,Z),(0,A',Z')]_{A^*\Gamma}\cr
=&(0,-[K(A),K(A')],0)\cr}\eqno(4.3)$$
for all smooth maps 
$A, A':U \longrightarrow \fg^*$,
$Z, Z':U \longrightarrow \fh$.
\endproclaim 

\demo
{Proof} The calculation will be postponed to the appendix.
\pf
\enddemo

Using $K$, we define

$${\Cal K}:A^*\Gamma \longrightarrow A\Gamma, 
(0_q,A,Z) \mapsto (0_q,K(A),0),\eqno(4.4)$$
and set ${\Cal R}^{\pm}={\Cal R}\pm {\Cal K}, R^{\pm}(q)=R(q)\pm K.$

\proclaim
{Proposition 4.2} 
\smallskip
\noindent (a) ${\Cal R}^{\pm}$ are morphisms of transitive Lie algebroids
and, as morphisms of vector bundles over $U$, are of locally constant
rank.  In particular,
$$[{\Cal R}^{\pm}(0,A,Z), {\Cal R}^{\pm}(0,A',Z')]_{A\Gamma}
={\Cal R}^{\pm}[(0,A,Z),(0,A',Z')]_{A^*\Gamma} \eqno(4.5)$$
for all smooth maps $A, A':U \longrightarrow \fg^*$,
$Z, Z':U \longrightarrow \fh.$  Moreover, ${\Cal R}^{\pm}$ are 
$H$-equivariant, where $H$ acts on $A^{*}\Gamma$ via
$h\cdot(0_q,A,Z) = (0_{Ad^{*}_{h^{-1}}} q, Ad^{*}_{h^{-1}} A, Ad_{h} Z)$
and the $H$-action on $A\Gamma$ is given by (3.4).
\smallskip
\noindent (b) $Im {\Cal R}^{\pm}$ are transitive Lie subalgebroids of 
$A\Gamma.$
\endproclaim

\demo
{Proof} (a) If $a$ and $a_*$ are the anchor maps of the Lie algebroids
$A\Gamma$ and $A^*\Gamma$ , it is easy to check that they are surjective
submersions which satisfy $a\circ {\Cal R}^{\pm}
=a_*$.  On the other hand, it follows from Lemma 4.1 that (4.5)
holds if and only if
$$\eqalign{& {\Cal K} [(0,A,Z),(0,A',Z')]_{A^*\Gamma}\cr
     =\,& [{\Cal R}(0,A,Z), {\Cal K}(0,A',Z')]_{A\Gamma}
         + [{\Cal K}(0,A,Z), {\Cal R}(0,A',Z')]_{A\Gamma}.\cr}\eqno(4.6)$$
Now, for $q\in U$, we have
$$
\aligned
       &{\Cal K} [(0,A,Z),(0,A',Z')]_{A^*\Gamma} (q)\\
      =&(0_q, K \bigl(dA'(q)(\iota^{*}A(q)-ad^{*}_{Z(q)} q)
         -ad^{*}_{R(q)A(q)-Z(q)}A'(q))  \\
       \,\,\,& -(A\leftrightarrow A',Z\leftrightarrow Z'),0 \bigr)
\endaligned
$$
where $(A\leftrightarrow A',Z\leftrightarrow Z')$ denote terms
which can be obtained from the previous ones by interchanging
$A$ and $A'$, $Z$ and $Z'$.
On the other hand, 
$$
\aligned
       &[{\Cal R}(0,A,Z),{\Cal K}(0,A',Z')]_{A\Gamma} (q) \\
     = &(0_q, K(dA'(q)(\iota^{*}A(q)-ad^{*}_{Z(q)} q))
         +[-\iota Z(q)+R(q)A(q),K(A'(q))],0)
\endaligned
$$ and similarly for $-[{\Cal R}(0,A',Z'),{\Cal K}(0,A,Z)]_{A\Gamma} (q).$
From these formulas, it follows that (4.6) holds if and only if
$$
\aligned
       &K(ad^{*}_{R(q)A(q)-Z(q)}A'(q))+[-\iota Z(q) + R(q)A(q), K(A'(q))]\\
       & \, -(A \leftrightarrow A', Z\leftrightarrow Z')=0.
\endaligned
$$
But the latter follows from the $G$-equivariance of $K$ and this proves
the first part of the assertion.(The fact that ${\Cal R}^{\pm}$
are of locally constant rank follows from Theorem 1.6 on page 190 of 
\c{M}.)   To show that ${\Cal R}^{\pm}$ are
$H$-equivariant, note that by definition,
$$
\aligned
       & {\Cal R}^{\pm} (h\cdot(0_q,A,Z))\\
     =\, &\bigl(0_{Ad^{*}_{h^{-1}} q}, -Ad_{h}Z + R^{\pm} (Ad^{*}_{h^{-1}}q)
         Ad^{*}_{h^{-1}} A,
         \iota^{*}Ad^{*}_{h^{-1}}A - ad^{*}_{Ad_{h}Z} Ad^{*}_{h^{-1}}q\bigr).
\endaligned
$$
But from the $H$-equivariance of $R$ and $K$, we have
$R^{\pm}(Ad^{*}_{h^{-1}}q) Ad^{*}_{h^{-1}}A = Ad_{h} R^{\pm}(q)A.$
On the other hand, it is straightforward to check that
$ad^{*}_{Ad_{h}Z} Ad^{*}_{h^{-1}}q = Ad^{*}_{h^{-1}} ad^{*}_{Z} q.$
Substituting into the above expression for
${\Cal R}^{\pm} (h\cdot(0_q,A,Z))$, the desired conclusion follows.
\newline
(b) This is a consequence of (a).
\pf
\enddemo

In the rest of the section, we shall assume $\fg$ has an ad-invariant 
non-degenerate pairing $(\cdot, \cdot)$ such that
$(\cdot, \cdot)|_{\fh \times \fh}$ is also non-degenerate.
Without loss of generality,
we shall take the map $K:\fg^* \longrightarrow \fg$ in the above
discussion to be the identification map induced by $(\cdot, \cdot)$.
Indeed, with the identifications $\fg^* \simeq \fg$, 
$\fh^* \simeq \fh$, we shall regard
$R(q)$ as taking values in $End (\fg)$, and the left and right gradients
as well as the dual maps are computed using $(\cdot, \cdot)$.  Also,
we have $ad^* \simeq -ad$, $\iota^* \simeq \Pi_{\fh}$, where $\Pi_{\fh}$
is the projection map to $\fh$ relative to the direct sum decomposition
$\fg = \fh \oplus \fh^{\perp}$.  We shall keep, however, the notation
$A^*\Gamma$ although as a set it can be identified with $A\Gamma.$

We now introduce the following subbundles of the adjoint bundle
$Ker\,a = \bigl \{ (0_q,X,0)\mid q\in U, X\in \fg \bigr \}$ of \, $A\Gamma$:
$${\Cal I}^{+} =\bigl \{(0_q,X,0)\in Ker\,a\mid q\in U,
   {\Cal R}^{-} (0_q,X,Z)=0 \,\,\,\hbox{for some}\,\,\, Z \in \fh \bigr \},
\eqno(4.7a)$$
$${\Cal I}^{-} =\bigl \{(0_q,X,0)\in Ker\,a\mid q\in U,
   {\Cal R}^{+} (0_q,X,Z)=0\,\,\, \hbox{for some} \,\,\, Z \in \fh \bigr \}.
\eqno(4.7b)$$

\proclaim
{Proposition 4.3} ${\Cal I}^{\pm}$ are ideals of the transitive Lie
algebroids $Im {\Cal R}^{\pm}$.
\endproclaim

\demo
{Proof} We shall prove the assertion for ${\Cal I}^+.$  First of all, it
is easy to show that ${\Cal I}^+ \subset Im {\Cal R}^{+}.$  
Let $(0,-\iota Z+R^{+}X, \Pi_{\fh}X + ad_{Z}(\cdot))\in 
Sect(U, Im {\Cal R}^{+})$ and $(0,X',0)\in Sect (U, {\Cal I}^+)$,
where $Z:U\longrightarrow \fh$, $X,X':U\longrightarrow \fg$ are
smooth maps.  From the expression for $[\cdot,\cdot]_{A\Gamma}$,
we have
$$\eqalign{&[(0,-\iota Z+R^{+}X, \Pi_{\fh}X + ad_{Z}(\cdot)),
  (0,X',0)]_{A\Gamma} (q)\cr
   =\,& (0_q,dX'(q)(\Pi_{\fh} X(q) + ad_{Z(q)} q) +
       [-\iota Z(q)+R^{+}(q)X(q),X'(q)],0)\cr}\eqno(4.8)$$
for $q\in U$. This shows
$$[(0,-\iota Z+R^{+}X, \Pi_{\fh}X + ad_{Z}(\cdot)),
  (0,X',0)]_{A\Gamma}\in Sect (U,Ker\,a).$$
On the other hand, from the assumption that
$(0,X',0)\in Sect(U,{\Cal I}^{+})$, we must have
${\Cal R}^{-}(0_q,X'(q),Z'(q))=0$ for some smooth map
$Z': U\longrightarrow \fh.$  Hence we obtain 
$(0,X',0)={\Cal R}^{+}(0,X'/2,Z'/2).$  Therefore, on using Proposition
4.2 (a), it follows that
$$\eqalign{& [(0,-\iota Z+R^{+}X, \Pi_{\fh}X + ad_{Z}(\cdot)),
             (0,X',0)]_{A\Gamma} (q)\cr
          =\,& {\Cal R}^{+} [(0,X,Z), (0,X'/2,Z'/2)]_{A^{*}\Gamma} (q)\cr
          =\,& {\Cal R}^{+} (0_q,X''(q),Z''(q))\cr
          =\,& (0_q,-\iota Z''(q)+R^{+}(q)X''(q),\Pi_{\fh} X''(q)
             +[Z''(q),q])\cr}\eqno(4.9)$$
where by (2.7), we find
$$\eqalign{X''(q)=\,&{1\over 2} dX'(q)(\Pi_{\fh}X(q) + ad_{Z(q)} q)
             +{1 \over 2} [R(q)X(q)-Z(q),X'(q)]\cr
            &  +{1 \over 2} [X(q), R(q)X'(q)-Z'(q)]\cr}\eqno(4.10)$$
and
$$\eqalign{Z''(q)=\,&{1\over 2} dZ'(q)(\Pi_{\fh}X(q) + ad_{Z(q)} q)
            -{1\over 2}[Z,Z'](q)\cr
           &  + {1\over 2} (dR(q)(\cdot)X(q),X'(q)).
            \cr}\eqno(4.11)$$
By equating the last components of the expressions in (4.8) and (4.9), 
we have
$$\Pi_{\fh} X''(q) + [Z''(q),q] = 0. \eqno(4.12)$$
Similarly, by equating the second components of the expressions in
(4.8) and (4.9),
we find
$$\eqalign{& dX'(q)(\Pi_{\fh} X(q) + ad_{Z(q)} q) +
       [-\iota Z(q)+R^{+}(q)X(q),X'(q)]\cr
      =&-\iota Z''(q)+R^{+}(q)X''(q)\cr}\eqno(4.13)$$
Now, from the relation ${\Cal R}^{-}(0_q,X'(q),Z'(q))=0$, it follows
that (4.10) can be rewritten as
$$\eqalign{&dX'(q)(\Pi_{\fh} X(q) + ad_{Z(q)} q) +
       [-\iota Z(q)+R^{+}(q)X(q),X'(q)]\cr
       =\, & 2 X''(q).\cr} \eqno(4.14)$$ 
Substitute this into (4.12), we obtain
$$\eqalign{\Pi_{\fh} & \{dX'(q)(\Pi_{\fh} X(q) + ad_{Z(q)} q) +
       [-\iota Z(q)+R^{+}(q)X(q),X'(q)]\}\cr
       & +[2Z''(q),q]=0.\cr}$$
Next, (4.13) and (4.14) yield
$$
\aligned
       &R^{-}(q)\{dX'(q)(\Pi_{\fh} X(q) + ad_{Z(q)} q) +
        [-\iota Z(q)+R^{+}(q)X(q),X'(q)]\}\\
      =\, & 2Z''(q).
\endaligned
$$
From the last two relations, we can now conclude that
$$[(0,-\iota Z+R^{+}X, \Pi_{\fh}X + ad_{Z}(\cdot),
  (0,X',0)]_{A\Gamma}\in Sect (U, {\Cal I}^{+}),$$
as desired.
\pf
\enddemo

Consider now the quotient vector bundles $Im {\Cal R}^{\pm}/{\Cal I}^{\pm}$
equipped with the quotient transitive Lie algebroid structures.

\proclaim
{Proposition 4.4} The map $\theta : Im {\Cal R}^{+}/{\Cal I}^{+}
\longrightarrow Im {\Cal R}^{-}/{\Cal I}^{-}$ defined by
$$\theta ({\Cal R}^{+} (0_q,X,Z) + {\Cal I}^{+}_q)=
  {\Cal R}^{-} (0_q,X,Z) + {\Cal I}^{-}_q$$
is an isomorphism of transitive Lie algebroids.
\endproclaim

\demo
{Proof} We first show that $\theta$ is well-defined.  To do so, 
suppose ${\Cal R}^{+} (0_q,X,Z) \equiv {\Cal R}^{+} (0_q,X',Z')$
(mod ${\Cal I}^{+}_q$).  Then there exists $Z''\in \fh$ such that
$$\Pi_{\fh} (X-X') + [Z-Z',q] = 0,$$
$$R^{-}(q)(-\iota (Z-Z') + R^{+}(q) (X-X')) =\iota Z'',$$
and
$$-(Z-Z') + \Pi_{\fh} (R^{+}(q)(X-X')) = ad_{q} Z''.$$
From these equations, we infer that
$$-\iota (Z-Z') + \Pi_{\fh} (R^{-}(q) (X-X')) = ad_{q} (Z''-2(Z-Z')),$$
and
$$R^{+}(q)(-\iota (Z-Z') +R^{-}(q)(X-X'))= Z''-2(Z-Z').$$
Hence we have ${\Cal R}^{-} (0_q,X,Z) \equiv {\Cal R}^{-} (0_q,X',Z')$
(mod ${\Cal I}^{-}_q$), as desired.  We shall skip the argument to
show that $\theta$
is 1:1 as it is  similar to the one above. The proof of the proposition
is therefore complete (it being clear that $\theta$ is a morphism
by Proposition 4.2 (a)). 
\pf
\enddemo

To formulate our next result,introduce the Lie algebroid direct sum 
$A\Gamma {\underset TU \to \oplus} A\Gamma.$
Clearly, this is the Lie algebroid of the product groupoid
$P = \Gamma {\underset U\times U\to \times} \Gamma \rightrightarrows U$ 
($\simeq$ the trivial Lie groupoid $U \times (G \times G)\times U$).
For later usage, we shall denote the structure maps (target, source etc.)
of $P$ by $\alpha_{P}$, $\beta_{P}$, and so forth.

\proclaim
{Theorem 4.5} (a)The map $({\Cal R}^{+}, {\Cal R}^{-}):A^{*}\Gamma
\longrightarrow A\Gamma {\underset TU \to \oplus} A\Gamma$ is a monomorphism
of transitive Lie algebroids.  In particular, the coboundary dynamical Lie
algebroid $(A^{*}\Gamma, [\cdot, \cdot]_{A^{*}\Gamma})$ is integrable.
\smallskip
\noindent (b)  $Im ({\Cal R}^{+}, {\Cal R}^{-})$ is the Lie subalgebroid
$$ \bigl \{({\Cal X}_+,{\Cal X}_-)\in (Im{\Cal R}^{+}
      {\underset TU \to\oplus}Im{\Cal R}^{-})_{q}\mid q\in U,\, 
      \theta({\Cal X}_{+}+{\Cal I}^{+}_q) = 
      {\Cal X}_{-} + {\Cal I}^{-}_q \bigr \} \eqno(4.15) 
$$
of  $Im{\Cal R}^{+} {\underset TU \to\oplus}Im{\Cal R}^{-}$.
\endproclaim

\demo
{Proof} (a) See \c{L1} for the proof.
\newline
(b) Denote by $A\Gamma_{R}$ the subbundle of 
$Im{\Cal R}^{+} {\underset TU \to\oplus}Im{\Cal R}^{-}$
defined in (4.15). It is clear that
$Im ({\Cal R}^+,{\Cal R}^-) \subset A\Gamma_{R}.$  Conversely,
suppose $((0_{q},X_{+},Z),(0_{q},X_{-},Z))\in A\Gamma_{R}$.  Then
there exist
$(0_q,X,\widetilde Z)$, $(0_q,X',\widetilde Z')\in A\Gamma$ such
that $(0_q,X_{+},Z) ={\Cal R}^{+} (0_q,X,\widetilde Z)$
and $(0_q,X_{-},Z) = {\Cal R}^{-} (0_q,X',\widetilde Z').$  Moreover,
from the property that 
$\theta ((0_q,X_{+},Z) + {\Cal I}^{+}_q) = (0_q,X_{-},Z) + {\Cal I}^{-}_q,$
we find
${\Cal R}^{-}(0_q,X-X',\widetilde Z -\widetilde Z')\equiv 0$ 
(mod ${\Cal I}^{-}_{q}$).  Let
$X'' = -\iota(\widetilde Z -\widetilde Z') + R^{-}(q) (X-X')$.
Then it follows from the definition of ${\Cal I}^{-}_q$ that there
exists $Z''\in \fh$ such that
${\Cal R}^{+}(0_q,X'',Z'') = 0.$  Now, consider the element
$(0_q,X + {1\over 2} X'', \widetilde Z + {1\over 2} Z'')\in A\Gamma.$
Clearly, 
${\Cal R}^{+}(0_q,X + {1\over 2} X'', \widetilde Z + {1\over 2} Z'')
= (0_q,X_{+},Z)$.  On the other hand,
$$\eqalign{& 
{\Cal R}^{-} (0_q,X + {1\over 2} X'', \widetilde Z + {1\over 2} Z'')\cr
= & (0_q,X_{-}, Z) + (0_q,X'',0) +{1\over 2} {\Cal R}^{-}(0_q,X'',Z'').\cr}$$
But as 
$$
\aligned
       &{\Cal R}^{-}(0_q,X'',Z'') \\
     = &{\Cal R}^{+}(0_q,X'',Z'')-(0_q,2X'',0)\\
     = &-(0_q,2X'',0),
\endaligned
$$
it follows from the above that 
${\Cal R}^{-} (0_q,X + {1\over 2} X'', \widetilde Z + {1\over 2} Z'')=
(0_q,X_{-}, Z).$  Thus we have shown that
$((0_{q},X_{+},Z),(0_{q},X_{-},Z))\in Im ({\Cal R}^{+}, {\Cal R}^{-}).$
\pf
\enddemo

The connection between (mDYBE) and our factorization theory is contained
in the decomposition
$$\eqalign {(0_q,X,0)
             &={1 \over 2}{\Cal R}^+ (0_q,X,0)\cr
             &-{1 \over 2}{\Cal R}^- (0_q,X,0)\cr}\eqno(4.16)
$$
where the element $(0_q,X,0)$ on the left hand side of (4.16) is in 
the adjoint bundle $Ker\,a$ of $A\Gamma.$ 
The reader should 
note that  the vector bundles
$\bigl \{{\Cal R}^{\pm} (0_q,X,0)\mid q\in U,\,X\in \fg \bigr \}$
are not Lie subalgebroids of $A\Gamma$ unless $R$ is
a constant r-matrix.  As we pointed out in \c{L1}, this fact has repercussion
when we try to formulate a global version of the decomposition in (4.16)
(see Corollary 4.6 below). 
 
In the rest of the section, we shall assume both $G$ and $U$ are 
simply-connected. Let $\Gamma^*$ be the unique source-simply connected 
Lie groupoid
which integrates $(A^{*}\Gamma,$ 
$[\cdot, \cdot]_{A^{*}\Gamma})$.  
Then $({\Cal R}^{+}, {\Cal R}^{-})$ can be lifted up to a unique
monomorphism of Lie groupoids $\Gamma^{*}\longrightarrow\Gamma
{\underset U \times U \to \times}\Gamma$ which we shall denote
by the same symbol.  Now, denote by 
${\Cal I}\Gamma = \{\,(u,g,u)\mid u\in U, g\in G\,\}$  the gauge
group bundle of $\Gamma$.  We let
$j: \Gamma {\underset U \times U \to \times}
\Gamma \longrightarrow {\Cal I}\Gamma$ be the map defined by
$j(a,b) = ab^{-1}$ and let ${\widetilde  m} = j \circ 
({\Cal R}^{+}, {\Cal R}^{-})$.

For the sake of completeness, we include the following Corollary of 
Theorem 4.5 (a) which (essentially) gives a global version of the 
decomposition in (4.16) which we mentioned above (the reader can
find the proof in \c{L1}).  For its formulation, note that the Lie groupoid
of $\{(0_{q}, 0, Z)\mid q\in U, Z\in \fh\}\subset A^{*}\Gamma$ is
$H \times U$, with target and source maps $\alpha^{\prime} (h,u) =u$,
$\beta^{\prime} (h,u)= Ad_{h} u$ and multiplication map
$m^{\prime} ((h,u),(k,Ad_{h} u))=(kh,u)$ (this is isomorphic
to the Hamiltonian unit in \c{LP2}).  On the other hand, the Lie groupoid of
${\Cal R}^{\pm} \bigl \{(0_{q},0,Z)\mid q\in U, Z\in \fh \bigr \}$
is given by  $E =\{\,(u, h, Ad_{h^{-1}}u) \mid u\in U,\,
h\in H\,\}$ and ${\Cal R}^{\pm}$ embeds $H\times U$ in $E$,
${\Cal R}^{\pm}\mid H\times U: (h,u)\mapsto (u, h^{-1}, Ad_{h} u).$ 
Clearly, the diagonal $\Delta(E)$ of  
$E {\underset U \times U \to \times} E$ acts
on $Im ({\Cal R}^{+}, {\Cal R}^{-})$ from the right via the
simple formula 
$$
\aligned
       &((u, k_{+}, v),(u, k_{-},v)).((v,h,Ad_{h^{-1}}v),(v, h, Ad_{h^{-1}}v))
        \\ 
   =& ((u, k_{+} h, Ad_{h^{-1}}v), (u,k_{-} h,Ad_{h^{-1}}v))
\endaligned
$$ 
and the map
$j \mid Im ({\Cal R}^{+}, {\Cal R}^{-})$ is constant on the orbits
of this action.

\proclaim
{Corollary 4.6} Suppose $U$ is simply-connected, then 
$j \mid Im ({\Cal R}^{+}, {\Cal R}^{-})$
induces a one-to-one map
${\widehat j}: {Im ({\Cal R}^{+}, {\Cal R}^{-})}/
\Delta(E) \longrightarrow
{\Cal I}\Gamma$.  Therefore, for each $\gamma \in Im \,{\widetilde m}$,
there exists unique $[\,(\gamma_+,\gamma_-)\,]$ in the homogeneous
space ${Im ({\Cal R}^{+}, {\Cal R}^{-})} / \Delta(E)$
such that ${\widehat j} ([\,(\gamma_+,\gamma_-)\,]) =\gamma$.
\endproclaim

Let $f\in I(\fg)$ and consider the Hamilton's equation generated by
$F = Pr^{*}_{3} f$. Then according to
Corollary 2.7 (a), we can express its restriction to the invariant manifold 
$\gamma^{-1}(0) = U \times \{0\}\times \fg$  in the
form
$$\eqalign{& {d \over dt}\, (q,0,X) \cr
  =\, & (\,\Pi_{\fh}\, df(X), 0, [X, R(q)df(X)]\,).\cr}\eqno(4.17)$$

In the next theorem, we shall express the solution of (4.17)
using the adjoint representation of $\Gamma$ on its adjoint bundle
$Ker\,a$, defined by ${\bA}_{\gamma} (q,0,X) = (q^{\prime},0,
Ad_{k}X)$, for $\gamma = (q^{\prime},k,q)\in \Gamma$.  
We shall also make the identifications
$A\Gamma$, $A^{*}\Gamma \simeq U\times \fh\times \fg$ throughout.  Thus the
element $(0,0,df(X_{0}))$ which appears in the theorem below
will denote the constant section of $Ker\,a$ such that
$(0,0,df(X_{0})) (q) = (q, 0, df(X_{0}))$ for $q\in U$.

\proclaim
{Theorem 4.7} Suppose that $f\in I(\fg)$, $F=Pr^{*}_{3} f$ and $q_0\in U$,
where $U$ is simply connected.  Then for some 
$0 < T \leq \infty$, there exists a unique element 
$(\gamma_{+}(t),\gamma_{-}(t))=
((q_0, k_{+} (t), q(t)), (q_0, k_{-} (t), q(t))) \in
Im ({\Cal R}^{+}, {\Cal R}^{-})$
for $0 \leq t <T$ which is smooth in t, solves the factorization
problem

$$exp \{2\, t(0,0,df(X_0))\}(q_0)
      =\,\gamma_{+} (t)\,\gamma_{-} (t)^{-1}\eqno(4.18)
$$
and satisfies 

$$\eqalign {(T_{\gamma_{+} (t)} {\bbl}_{{\gamma_{+}(t)}^{-1}} {\dot\gamma_{+}(t)},
T_{\gamma_{-} (t)}{\bbl}_{{\gamma_{-}(t)}^{-1}}{\dot \gamma_{-}(t)})
\in \, &
({\Cal R}^+,{\Cal R}^-) (\{q(t)\}\times\{0\}\times\fg)\cr}
\eqno(4.19a)$$
with
$$\gamma_{\pm}(0) = (q_{0},1,q_{0}).\eqno(4.19b)$$

\noindent Moreover, the solution of (4.17) with initial data
$(q,0,X)(0)= (q_0,0,X_0)$ (i.e. the induced flow on 
$\gamma^{-1}(0)$ generated by $F$) is given by the formula
$$(q(t),0,X(t)) ={\bA}_{{\gamma_{\pm}(t)^{-1}}} (q_0,0,X_0).\eqno(4.20)$$
\endproclaim

\demo
{Proof}  The uniqueness of the element $(\gamma_{+} (t),\gamma_{-}
(t))$ is proved in the same way as in \c{L1} and makes crucial use
of Corollary 4.6.  

Assuming the existence of the factors for the moment, we claim
that $(q(t),0,X(t))$ as given by (4.20) solves (4.17).  
First of all, we have 
$$
\aligned
       & {\bA}_{{\gamma_{+}(t)}^{-1}} (q_0,0,X_0) \\
    =\,& (q(t), 0, Ad_{k_{+}(t)^{-1}} X_0) \\
    =\,& (q(t), 0, Ad_{k_{-}(t)^{-1}} Ad_{e^{-2\,tdf(X_0)}} X_0)\\
    =\,& (q(t), 0, Ad_{k_{-}(t)^{-1}} X_0)\\
    =\,& {\bA}_{{\gamma_{-}(t)}^{-1}} (q_0,0,X_0)
\endaligned
$$
where we have used the fact that $[\,df(X_0), X_0\,] = 0$.
Take
$$ (q(t), 0, X(t)) = {\bA}_{\gamma_{+}(t)^{-1}} (q_0, 0, X_0).$$
By differentiating the expression, we have
$$
\aligned
       & {d\over dt}\, (q(t), 0, X(t))\\
    =\,& (\dot q(t), 0, \bigl[\,X(t), T_{k_{+}(t)} l_{{k_{+}(t)}^{-1}} 
       \dot k_{+}(t) \,\bigr]).
 \qquad \qquad (*)
\endaligned
$$
On the other hand, by rewriting (4.18) in the form
$$exp\{2t(0,df(X_0),0)\} (q_0)\,\,\gamma_{-}(t) =\gamma_{+} (t),$$
we have, upon differentiation, that 
$$
\aligned
       & T_{\gamma_{+} (t)} {\bbl}_{{\gamma_{+}(t)}^{-1}} {\dot\gamma_{+}(t)}
        - T_{\gamma_{-} (t)}{\bbl}_{{\gamma_{-}(t)}^{-1}}{\dot \gamma_{-}(t)}\\
    =\,&2\, {\bA}_{{\gamma_{-}(t)}^{-1}} (q_0, 0, df(X_0)).
\endaligned
$$
But
$$
\aligned
       & {\bA}
_{\gamma_{-}(t)^{-1}} (q_0,0,df(X_0))\\
     =\, & (q(t), 0, Ad_{k_{-}(t)^{-1}} df(X_0)) \\
     =\, & (q(t), 0, df(X(t)))
\endaligned
$$
as $f\in I(\fg)$.  Hence it follows that
$$
\aligned
       &T_{\gamma_{+} (t)} {\bbl}_{{\gamma_{+}(t)}^{-1}} {\dot\gamma_{+}(t)}
        -T_{\gamma_{-} (t)}{\bbl}_{{\gamma_{-}(t)}^{-1}}{\dot \gamma_{-}(t)}\\
      =\,&2 (q(t), 0, df(X(t))).
\endaligned
$$
From the property of  $\gamma_{\pm}$ in (4.19), we can now conclude
that
$$T_{\gamma_{\pm} (t)}{\bbl}_{{\gamma_{\pm}(t)}^{-1}} {\dot \gamma_{\pm}(t)} 
    = {\Cal R}^{\pm} (q(t), 0, df(X(t))).$$
But
$$T_{\gamma_{+} (t)} {\bbl}_{{\gamma_{+}(t)}^{-1}} {\dot\gamma_{+}(t)}
     =\,(q(t), \dot q(t), T_{k_{+}(t)} l_{{k_{+}(t)}^{-1}} \dot k_{+}(t)),$$ 
while
$$ {\Cal R}^{+} (q(t), 0, df(X(t)))
    =\, (q(t), \Pi_{\fh}\,df(X(t)), R^{+} (q(t)) df(X(t))).$$
By equating the two expressions, we obtain
$$\dot q(t) = \Pi_{\fh}\,df(X(t)),$$
and
$$T_{k_{+}(t)} l_{{k_{+}(t)}^{-1}} \dot k_{+}(t)=
  R^{+} (q(t)) df(X(t)).$$
Therefore, on substituting into (*), we find
$$
\aligned
       & {d\over dt} (q(t), 0, X(t))\\
    =\,& (\Pi_{\fh}\,df(X(t)), 
           [\, X(t), R(q(t))df(X(t))\,]),
\endaligned
$$
as claimed.

To prove the existence of the factors $\gamma_{\pm}(t)$, simply
solve the initial value problems
$$\dot k_{\pm}(t) = T_{e}l_{k_{\pm}(t)} R^{\pm}(q(t)) 
  df(X(t)), \qquad  k_{\pm} (0) = 1, \qquad\quad (**)$$
where $q(t)$, $X(t)$ are the solutions of (4.17) with
initial data $(q, 0,X)(0) = (q_0,0,X_0)$ (which are known
to exist by ODE theory). Set 
$\gamma_{\pm} (t) = (q_0, k_{\pm}(t), q(t))$. As can be easily verified, we
can combine the equations for $q(t)$, $k_{\pm}(t)$ into one
single equation for 
$(\gamma_{+}(t),\gamma_{-}(t))$:
$$
\aligned
       &{d \over dt} (\gamma_{+}(t), \gamma_{-}(t)) \\
    =\,&\bigl(T_{\epsilon(q(t))} \bbl_{\gamma_{+}(t)}
         {\Cal R}^{+} (q(t), 0, df(X(t))),
         T_{\epsilon(q(t))} \bbl_{\gamma_{-}(t)}
         {\Cal R}^{-} (q(t), 0, df(X(t)))\bigr) \\
    =\,&T_{\epsilon_{P} (\beta_{P} (\gamma_{+}(t), \gamma_{-}(t)))}\,
        \bbl^{P}_{(\gamma_{+}(t), \gamma_{-}(t))}
        ({\Cal R}^{+},{\Cal R}^{-}) (q(t), 0,df(X(t)))
     \quad \quad (***)
\endaligned
$$
where          
$\bbl^{P}_{ (\gamma_{+}(t), \gamma_{-}(t))}$ represents left translation
by $(\gamma_{+} (t), \gamma_{-}(t))$ in the product groupoid
$P = \Gamma {\underset U\times U\to \times} \Gamma \rightrightarrows U$. 
Clearly, what we have just written down is a well-defined equation
for $(\gamma_{+}(t),\gamma_{-}(t))\in Im ({\Cal R}^{+},{\Cal R}^{-}).$
Moreover, from the initial conditions for $k_{\pm}(t)$ and $q(t)$, we
have $(\gamma_{+}(0),\gamma_{-}(0)) \in Im ({\Cal R}^{+},{\Cal R}^{-}).$

Now, from the equations for $k_{\pm}$ in (**),  we find
$$
\aligned
       &T_{\gamma_{+}(t)\gamma_{-}(t)^{-1}} \bbl_{(\gamma_{+}(t)\gamma_{-}
        (t)^{-1})^{-1}}{d \over dt} \gamma_{+}(t)\, \gamma_{-}(t)^{-1}  \\
    =\,& (q_0, 0, 2\, df (Ad_{k_{-}(t)} X(t))).
\endaligned
$$
But from the equation for $k_{-}(t)$ and $X(t)$
, we have
$$
\aligned
       & {d \over dt} Ad_{k_{-}(t)} X(t) \\
   =\, & Ad_{k_{-}(t)} \dot X(t) + [\, T_{k_{-}(t)} 
         r_{{k_{-}(t)}^{-1}} \dot k_{-}(t), Ad_{k_{-}(t)} X(t)\,]\\
   =\, & [\,Ad_{k_{-}(t)} X(t), Ad_{k_{-}(t)} R(q(t))df(X(t))\,]\\
     & + [\,Ad_{k_{-}(t)} R^{-}(q(t))df(X(t)), Ad_{k_{-}(t)} X(t)
         \,]\\
   =\, & 0.
\endaligned
$$
Therefore, $Ad_{k_{-}(t)} X(t) = X_0$ and so
$$
\aligned
       &T_{\gamma_{+}(t)\gamma_{-}(t)^{-1}} \bbl_{(\gamma_{+}(t)\gamma_{-}
        (t)^{-1})^{-1}}{d \over dt} \gamma_{+}(t)\, \gamma_{-}(t)^{-1}  \\
    =\,& (q_0,0,2\, df (X_{0})).
\endaligned
$$ 
As $\gamma_{+}(t)\,\gamma_{-}(t)^{-1} = (q_{0}, k_{+}(t)k_{-}(t)^{-1}, q_{0}),$
this shows that $k_{+}(t)k_{-}(t)^{-1} = e^{2t df(X_{0})}$ and 
consequently, 
$$exp \{2\,t(0,df(X_0),0)\}(q_0) = \gamma_{+}(t)\, \gamma_{-}(t)^{-1}.$$
Thus it remains to show that condition (4.19 a) is satisfied.  But this is
immediate from (***).  This
completes the proof.
\pf
\enddemo

\proclaim
{Corollary 4.8} Let $\psi_{t}$ be the induced flow on $\gamma^{-1}(0)
= U\times \{0\}\times \fg$ as
defined in (4.20) and let $\phi_{t}$ be the Hamiltonian flow of 
${\Cal F} =L^{*} f$ on $X$, where $L = Pr_{3}\circ\rho$ for a realization
map $\rho:X \longrightarrow A\Gamma$ satisfying A1-A3.  If we can solve
for $\phi_{t} (x)$, $x \in J^{-1} (\mu)$ explicitly from the relation
$\rho(\phi_{t}) (x) = \psi_{t} (\rho (x))$, then the formula
$\phi^{red}_{t} \circ \pi_{\mu} = \pi_{\mu} \circ \phi_{t} \circ i_{\mu}$
gives an explicit expression for the flow of the reduced Hamiltonian
${\Cal F}_{\mu} = {\widehat \rho}^{*} {\overline f}$.
\endproclaim
\smallskip
\noindent {\bf Remark 4.9} (a) The reader should not feel uneasy about
the use of the equation (**) above (which involve the solutions $q(t)$
and $X(t)$) to show the existence of the factors $k_{\pm}(t)$, and
which are then used in turn to construct $q(t)$ and $X(t)$.  As the
reader will see in Section 6 below, knowledge of the existence of 
the factorization facilitates its construction.
\smallskip
\noindent (b) If we take $K = {1\over 2} id_{\fg}$, which is what we
will need in Section 6, then the factorization problem in (4.18) has to
be replaced by $exp \{t (0,0,df(X_0))\}(q_0) =\,\gamma_{+} (t)\,
\gamma_{-} (t)^{-1}$.  Otherwise, the solution formula is the same
as before.
\smallskip
\noindent (c) There is a similar method for solving the Hamiltonian flows
generated by natural invariant functions on the gauge group bundles
of cobundary dynamical Poisson groupoids.  We shall refer the reader
to \c{L1} for details.
\smallskip
\noindent (d) For the hyperbolic spin Calogero-Moser systems and
the spin Toda lattices which we introduce in the next section,
the assumption in Corollary 4.8 (namely, we can solve for 
$\phi_{t} (x)$, $x \in J^{-1} (\mu)$ explicitly from the relation
$\rho(\phi_{t}) (x) = \psi_{t} (\rho (x))$) are actually not 
satisfied in general.  As the reader will see, some special structure of
these equations still enables us to obtain the Hamiltonian flows
on $J^{-1}(\mu)$ from the induced flows on $\gamma^{-1}(0)$.
\smallskip
\noindent (e) Clearly, Theorem 4.7 also applies in the case when $R$ is a 
constant r-matrix.  However, it is  possible to formulate an analog of this
result using the fact that the vector bundles 
$\bigl \{{\Cal R}^{\pm} (0_q,X,0)\mid q\in U,\,X\in \fg \bigr \}$
are Lie subalgebroids of $A\Gamma$ in this case, but we provide
no details here.
\bigskip
\bigskip

\subhead
5. \ A family of hyperbolic spin Calogero-Moser systems
and  the spin \linebreak \phantom{fak}\,\,\, Toda lattices
\endsubhead

\bigskip

In \c{EV}, the authors classified solutions of (mDYBE) for pairs
$(\fg, \fh)$ of Lie algebras, where $\fg$ is simple, and 
$\fh \subset \fg$ is a Cartan subalgebra.  The purpose of this
section is to introduce a natural family of hyperbolic spin 
Calogero-Moser systems associated with these solutions as another
application of Proposition 4.2 (a).  Remarkably, these models
admit scaling limits, and the result is a family of Hamiltonian
systems which may be regarded as a spin generalization of the
Toda lattice.

Let us begin with some notation.  Let
$\fg=\fh \oplus \sum_{\alpha \in\Delta} \fg_{\alpha}$
the root space decomposition of the simple Lie algebra
$\fg$ and let $(\cdot,\cdot)$ denote its Killing form.
For each $\alpha \in \Delta$, denote by $H_{\alpha}$ the element
in $\fh$ which corresponds to $\alpha$ under the isomorphism
between $\fh$ and $\fh^*$ induced by the Killing form $(\cdot,\cdot)$.
We fix a simple system of roots  $\pi=\{\alpha_1,\cdots, \alpha_N\}$ 
and denote by
$\Delta^\pm$ 
the corresponding positive/negative
system.  For any positive root $\alpha \in \Delta^+$, we  choose
root vectors $e_{\alpha} \in \fg_{\alpha}$ and
$e_{-\alpha} \in \fg_{-\alpha}$ which are dual with respect to
$(\cdot, \cdot )$ so that $[e_{\alpha} , e_{-\alpha}]=H_{\alpha}$.  We
also fix an orthonormal basis $(x_i)_{1\le i\le N}$ of $\fh$.
Lastly, for a subset of simple roots $\pi^{\prime}\subset \pi$, we shall
denote the 
root span of $\pi^{\prime}$ by
$<\pi^{\prime}>\subset \Delta$ and set ${\overline
\pi^{\prime}}^{\pm}= \Delta^\pm \setminus <\pi^{\prime}>^\pm.$

For any subset $\pi^{\prime}\subset\pi$, we consider the following 
$H$-equivariant
solution of the (mDYBE) (with $K=
{1\over 2} id_{\fg}$):

$$R(q)X= - \sum_{\alpha\in \Delta} \phi_{\alpha} (q) X_{\alpha} e_{\alpha}
\eqno(5.1a)$$
where
$$\eqalign {&\phi_\alpha (q) = {1\over 2}\,\, {\hbox { for }} \alpha\in
{\overline 
\pi^{\prime}}^+,\quad\phi_\alpha (q) = - {1\over 2} \,\,{\hbox { for }} \alpha\in
{\overline 
\pi^{\prime}}^-\cr
&\phi_\alpha (q)= {1\over 2} \coth ({1\over2} (\alpha(q))\,\, {\hbox {
for }} \alpha\in <\pi^{\prime}>,\cr}\eqno(5.1b)$$
and $X_{\alpha}= (X, e_{-\alpha}), \quad \alpha\in \Delta.$
From now onwards, we shall assume $G$ and $H$ are simply-connected.

Consider now  the coboundary dynamical Lie algebroid
$A^{*}\Gamma$ 
which corresponds to this particular choice of $R$.  By Proposition 4.2 (a), 
we know that the  
associated bundle maps ${\Cal R}^{\pm}$ are morphisms of 
Lie algebroids, hence it follows that the dual maps
$({\Cal R}^{\pm})^*= -{\Cal R}^{\mp}$ are Poisson maps,
when the domain and target are equipped with the corresponding 
Lie-Poisson structures.  Note that 
the Lie-Poisson structure $\{\,\cdot,\cdot \,\}_{A^{*}\Gamma}$
on the dual bundle $A^*\Gamma
\simeq TU \times \fg$ 
of the trivial Lie algebroid $A\Gamma$ is a product structure, as is evident
from the expression in Remark 2.5.  Hence we have $H$-equivariant
realizations of 
$A^{*}\Gamma$ (the dual of the trivial Lie algebroid $A\Gamma$) in 
the dual vector vector bundle $A\Gamma$ of the dynamical Lie algebroid
$A^{*}\Gamma$.

To summarize, we have the following.

\proclaim
{Proposition 5.1} $({\Cal R}^{\pm})^*$ are $H$-equivariant Poisson maps, where
$H$ acts on $A^{*}\Gamma$, $A\Gamma \simeq TU\times \fg$ by acting
on the  factor $\fg$ by adjoint action.  
\endproclaim

To construct the spin Calogero-Moser system associated to the dynamical
r-matrix $R$ in (5.1), introduce the quadratic function
$$ Q(\xi)={1 \over 2} (\xi,\xi), \quad \xi \in \fg.\eqno(5.2)$$
We shall take $\rho= ({\Cal R}^{+})^*$ to be our realization map
(the other case with $({\Cal R}^{-})^*$ is similar) and
let $L=\Pr_{3}\circ \rho$, as in (3.3).  Then
the spin Calogero-Moser system associated to $R$ is the Hamiltonian
system on $A^{*}\Gamma \simeq TU\times \fg$ generated by the
Hamiltonian
$${\Cal H} (q,p,\xi)=L^{*}Q(q,p,\xi)\eqno(5.3)$$

Write $p=\sum_{i} p_{i} x_{i}$, \, \,$\xi=\sum_{i} \xi_{i} x_{i} +
\sum_{\alpha \in \Delta} \xi_{\alpha} e_{\alpha}$, then we have

\proclaim
{Proposition 5.2} The Hamiltonian of the spin Calogero-Moser system
associated to the dynamical r-matrix $R$ in (5.1) is given by
$$\eqalign {{\Cal H} (q,p,\xi) =& {1 \over 2} \sum_{i} p_{i}^{2} +
         {1 \over 8} \sum_{i} \xi_{i}^{2} + {1 \over 2} \sum_{i}
         p_{i}\xi_{i} \cr
         &-{1 \over 8} \sum_{\alpha \in <\pi^{\prime}>} \frac{\xi_{\alpha}
         \xi_{-\alpha}} {sinh^{2} {1 \over 2} \alpha (q)}\cr}\eqno(5.4)
$$ 
and is invariant under the Hamiltonian $H$-action on 
$A^{*}\Gamma \simeq TU \times \fg$:
$$h\cdot(q,p,\xi)=(q,p,Ad_{h} \xi)\eqno(5.5)$$ 
with momentum map $J: TU\times \fg \longrightarrow \fh$ given by
$$ J(q,p, \xi) = - \Pi_{\fh}\, \xi.\eqno(5.6)$$
\endproclaim

Consider the level set $J^{-1}(0)$ which is invariant under the
flow $\phi_{t}$ generated by ${\Cal H}$.  Since $J=\gamma \circ \rho$,
where $\gamma$ is the momentum map in Proposition 3.1,we clearly have 
$\rho (J^{-1} (0)) \subset \gamma^{-1}(0)$.  Hence assumptions A1-A3
are satisfied.  Therefore, the family of functions 
$L^{*} I(\fg)$ Poisson commute on $J^{-1}(0)$ and hence descend to
Poisson commuting functions on the reduced Poisson variety
$J^{-1}(0)/H.$
 
\noindent{\bf Remark 5.3}  Note that if we consider the realization
map $\rho^{-} = ({\Cal R}^{-})^*= -{\Cal R}^{+}$  instead, then we would
have the slightly different Hamiltonian
$$\eqalign {{\Cal H}^{-} (q,p,\xi) =& {1 \over 2} \sum_{i} p_{i}^{2} +
         {1 \over 8} \sum_{i} \xi_{i}^{2} -{1 \over 2} \sum_{i}
         p_{i}\xi_{i} \cr
         &-{1 \over 8} \sum_{\alpha \in <\pi^{\prime}>} \frac{\xi_{\alpha}
         \xi_{-\alpha}} {\sinh^{2} {1 \over 2} \alpha (q)}\cr}
$$ 
and the associated Lax operator in this case is given by
$L^{-}(q,p,\xi) = p - R^{+} (q) \xi$.

\proclaim
{Proposition 5.4} The Hamiltonian equations of motion generated by
${\Cal H}$ on $A^{*}\Gamma$ are given by 
$$\eqalign{& \dot q = p + {1\over 2} \Pi_{\fh}\, \xi,\cr
           & \dot p = -{1\over 8} \sum_{\alpha \in <\pi^{\prime}>}
             \frac{\coth {1\over 2} \alpha(q)}
             {\sinh^{2} {1\over 2} \alpha (q)}
             \xi_{\alpha} \xi_{-\alpha} H_{\alpha},\cr
           & \dot \xi = \Bigl[\,\xi, {1\over 4} \Pi_{\fh}\, \xi
             + {1\over 2} p -{1\over 4} \sum_{\alpha \in <\pi^{\prime}>}
             \frac{\xi_{\alpha}} {\sinh^{2} {1\over 2} \alpha (q)}
             e_{\alpha}\,\Bigr]\cr
           & \,\,= [\,\xi, R^{+}(q) L(q,p,\xi)\,].\cr}\eqno(5.7)
$$
Moreover, under the  Hamiltonian flow, we have
$$\eqalign{& (\Pi_{\fh}\, \xi)^{\cdot} = 0 \cr
           & \dot L (q,p,\xi) = [\, L(q,p,\xi), R(q) L(q,p,\xi) \,]\cr
           &\qquad \qquad -dR(q)(\Pi_{\fh}\,\xi) L(q,p,\xi).}
           \eqno(5.8)$$
\endproclaim

\demo
{Proof} From the expression for the Poisson bracket in Remark 2.5, the
equations of motion are given by
 $\dot q = \delta_{2} {\Cal H}$, $\dot p = -\delta_{1} {\Cal H}$ and
$\dot \xi = [\,\xi, \delta {\Cal H}\,].$  Therefore, (5.7) follows by
a direct computation.  On the other hand, it follows from the definition
of $\rho$ that $m(q,p,\xi)=q$ and $\tau(q,p,\xi)= -\Pi_{\fh}\,\xi$ in
the notation introduced in (3.1)-(3.2).
Therefore, (5.8) is a consequence of Theorem 3.2 (c).
\pf
\enddemo

We shall solve Eqn.(5.7) on the level set $J^{-1}(0)$ 
(where $\Pi_{\fh}\,\xi = 0$) in Section 6 below.
In order to write down the equations of motion of the reduced
Hamiltonian system, we have to restrict to a smooth component of 
$J^{-1} (0)/H = U\times \fh \times (\fh^{\perp}/H)$.  For this purpose, we
consider 
 the following open submanifold of
$\fg$:
$${\Cal U} =\{\,\xi\in \fg \mid {\xi}_{\alpha_i} = (\xi, e_{-\alpha_i})
\neq 0, \quad i=1,\ldots, N \,\}. \eqno(5.9)$$
Clearly,  $TU\times {\Cal U}$ is a Poisson submanifold of 
$TU \times \fg \simeq A^{*}\Gamma$ and the $H$-action defined by (5.5)
induces a Hamiltonian action on $TU\times {\Cal U}$.  Therefore, 
the corresponding momentum map is given by the restriction of the
one in (5.6).  To simplify notation, we shall denote this
momentum map also by $J$ so that 
$J^{-1}(0) = TU \times (\fh^{\perp}\cap {\Cal U})$.

Now, recall from \c{LX2} that the formula
$$g(\xi ) =\exp{\Bigl(\sum_{i=1}^{N}
\sum_{j=1}^{N}(C_{ji} \log{\xi_{\alpha_{j}}})h_{\alpha_{i}}\Bigr)}\eqno(5.10)$$
defines an $H$-equivariant map $g: {\Cal U}\longrightarrow H$, where
$C= (C_{ij})$ is the inverse of the Cartan matrix and 
$h_{\alpha_{i}} = {2 \over (\alpha_i,\alpha_i)} H_{\alpha_{i}}$, 
$i=1,\ldots, N$.  Using $g$, we can identify the reduced space
$J^{-1}(0)/H = TU \times (\fh^{\perp} \cap {\Cal U}/H)$ with
$TU\times \fg_{red}$, where $\fg_{red}$ is the affine subspace
$\epsilon + \sum_{\alpha \in \Delta - \pi} {\Bbb C} e_{\alpha}$,
and $\epsilon = \sum_{j=1}^{N} e_{\alpha_{j}}$.  Indeed, if
we write $\alpha = \sum_{i=1}^{N} m^{i}_{\alpha} \alpha_{i}$ for 
each $\alpha\in \Delta$, then the identification map is given by
$$(q,p,[\xi]) \mapsto (q,p, Ad_{g(\xi)^{-1}} \xi),\eqno(5.11)$$
where explicitly,
$$ Ad_{g(\xi)^{-1}} \xi = \epsilon + \sum_{\alpha\in \Delta - \pi}
  \xi_{\alpha}\Bigl(\prod_{i=1}^{N} \xi_{\alpha_j}^{-m^{j}_{\alpha}}\Bigr)
  e_{\alpha}. \eqno(5.12)$$
Thus the natural projection 
$\pi_{0} : J^{-1}(0) \longrightarrow TU\times \fg_{red}$
is the map
$$(q,p, \xi)\mapsto  (q,p, Ad_{g(\xi)^{-1}} \xi).\eqno(5.13)$$

We shall write $s = \sum_{\alpha\in \Delta} s_{\alpha} e_{\alpha}$
for $s\in \fg_{red}$ (note that
$s_{\alpha_{j}} = 1\,\, \hbox{for}\, j= 1,\ldots, N$).
By Poisson reduction \c{MR}, the reduced manifold 
$TU\times \fg_{red}$ has a unique Poisson structure which
is a product structure , where the second factor
$\fg_{red}$ is equipped with the reduction (at 0) of the Lie-
Poisson structure on ${\Cal U}$ by the $H$-action.  Now the
symplectic leaves of $\fg_{red}$ are the symplectic reduction
of ${\Cal O} \cap {\Cal U}$ at 0, where ${\Cal O}\subset \fg$
is an adjoint orbit \c{MR}.  In other words, any symplectic
leaf of $\fg_{red}$ is of the form
$({\Cal O} \cap {\Cal U} \cap \fh^{\perp})/H$, and we shall
denote this by ${\Cal O}_{red}$.
Consequently, the symplectic leaves  of $TU \times \fg_{red}$ are
of the form $TU\times {\Cal O}_{red}$, which is of dimension
equal to $dim {\Cal O}$.
 Therefore, if ${\Cal H}$
is the Hamiltonian of the hyperbolic spin Calogero-Moser system in
(5.4), then its reduction ${\Cal H}_0$ on $TU \times \fg_{red}$ 
is given by 
$${\Cal H}_{0}(q,p,s) = {1 \over 2} \sum_{i} p_{i}^{2} -{1\over 4}
\sum_{\alpha \in <\pi^{\prime}>^{+}} \frac {s_{\alpha}
         s_{-\alpha}} {sinh^{2} {1 \over 2} \alpha (q)},\eqno(5.14)$$
where $s\in \fg_{red}$. 

\noindent {\bf Remark 5.5}  (a) In the special case where $\pi^{\prime}
=\pi$, the
Hamiltonian system generated by ${\Cal H}_{0}$ is isomorphic to the one
in Reshetikhin's paper \c{R}.
\smallskip
\noindent (b) The family of integrable hyperbolic spin Calogero-systems 
constructed
in this section are different from the ones in \c{LX2}.  Although they
look similar, however, their explicit integration requires different tools.  
To be more precise, the factorization
problems for the systems in \c{LX2} are associated with infinite dimensional
Lie groupoids whose vertex groups are loop groups.  The solution
of such factorization problems requires the use of algebraic geometry
(compare Section 6 and \c{L2}).  On the other hand, from the point of view
of proving complete integrability, the two distinct families
of hyperbolic systems also require totally different considerations.  We 
shall discuss these matters in subsequent publications.

\proclaim
{Proposition 5.6}  The Hamiltonian equations of motion generated by
${\Cal H}_0$ on the reduced Poisson manifold $TU\times \fg_{red}$
are given by
$$\aligned
         & \dot q = p, \\
         & \dot p = -{1\over 8} \sum_{\alpha \in <\pi^{\prime}>}
             \frac{\coth {1\over 2} \alpha(q)}
             {\sinh^{2} {1\over 2} \alpha (q)}
              s_{\alpha} s_{-\alpha} H_{\alpha}, \\
         & \dot s = [\,s, {\Cal M}\,] 
\endaligned
$$
where 
$${\Cal M} =  -{1\over 4} \sum_{\alpha \in <\pi^{\prime}>}
           \frac{s_{\alpha}} {\sinh^{2} {1\over 2} \alpha (q)} e_{\alpha}
          + {1\over 4} \sum_{i,j} C_{ji}
          \sum \Sb \alpha \in <\pi^{\prime}>-\pi^{\prime} \\
           \alpha_{j}-\alpha \in \Delta \endSb N_{\alpha, \alpha_{j}-\alpha} 
           \frac{s_{\alpha} s_{\alpha_{j}-\alpha}} 
           {\sinh^{2}{1\over 2} \alpha (q)}  
           h_{\alpha_i}.$$
(Here we use the notation $[e_{\alpha}, e_{\beta}] = N_{\alpha,\beta}
e_{\alpha +\beta}$ if $\alpha + \beta \in \Delta$.)
\endproclaim

\demo
{Proof} The first two equations are obvious from Proposition 5.4 and 
the definition of $s$.  To derive the equation of $s$, we
differentiate $s = Ad_{{g(\xi)}^{-1}} \xi$ with respect to $t$,
assuming $\xi$ satisfies the equation in Proposition 5.4 with
$\Pi_{\fh}\,\xi = 0$.  Then we have
$$\aligned
\dot s = & \Bigl[\,T_{g(\xi)^{-1}} r_{g(\xi)} {d\over dt} {g(\xi)}^{-1}, s\,
\Bigr ] + Ad_{{g(\xi)}^{-1}} \dot \xi\\
       =& \Bigl[\,s , {1\over 2} p -{1\over 4} \sum_{\alpha\in<\pi^{\prime}>}
        \frac{\xi_{\alpha}} {\sinh^{2} {1\over 2} \alpha (q)}
        e^{-\alpha(log g(\xi))} e_{\alpha}-
        T_{g(\xi)^{-1}} r_{g(\xi)} {d\over dt} {g(\xi)}^{-1}\,\Bigr].\qquad (*)
\endaligned
$$
By a direct computation, we find $\xi_{\alpha} e^{-\alpha(log\, g(\xi))} 
= \xi_{\alpha}(\prod_{i=1}^{N} \xi_{\alpha_j}^{-m^{j}_{\alpha}}) =s_{\alpha}$. 
 Meanwhile, by differentiating $g(\xi)^{-1}$, we obtain
$$-T_{g(\xi)^{-1}} r_{g(\xi)} {d\over dt} {g(\xi)}^{-1}=
   \sum_{i,j} C_{ji} {\dot \xi_{\alpha_j}} 
   {\xi_{\alpha_j}}^{-1} h_{\alpha_i}.$$
But 
$$\aligned
\dot \xi_{\alpha_j} &= (\dot \xi, e_{-\alpha_j})\\
                    &= \Bigl(\,\Bigl[\,\xi, {1\over 2} p -{1\over 4} 
                     \sum_{\alpha \in  <\pi^{\prime}>}
                   \frac{\xi_{\alpha}} {\sinh^{2} {1\over 2} \alpha (q)}
                    e_{\alpha}\,\Bigr], e_{-\alpha_j}\,\Bigr)\\
          &= -{1\over 2} \alpha_{j}(p)\xi_{\alpha_j} + {1\over 4} 
            \xi_{\alpha_j}\sum \Sb \alpha \in <\pi^{\prime}>-\pi^{\prime} \\
           \alpha_{j}-\alpha \in \Delta \endSb N_{\alpha, \alpha_{j}-\alpha} 
           \frac{s_{\alpha} s_{\alpha_{j}-\alpha}} 
           {\sinh^{2}{1\over 2} \alpha (q)}
\endaligned
$$
whereas ${1\over 2} p = \sum_{i,j} C_{ji} \alpha_{j}({1\over 2} p) 
h_{\alpha_{i}}$.
Therefore, on substituting the above expressions into (*), the
desired equation follows.
\pf
\enddemo

In the rest of the  section, we shall describe a scaling limit of the 
hyperbolic spin Calogero-Moser systems.  More precisely, we 
consider 
$$\eqalign{& q = x + 2 \tau w, \quad \tau > 0,\cr
          & \xi_i = \eta_i, \quad 1 \le i \le N,\cr
          & \xi_{\alpha} = \eta_{\alpha} \, e^{\tau},  \quad \alpha\in
            \Delta,\cr}\eqno(5.15)
$$
in the limit $\tau \to \infty$, where 
$$w = \sum_{\alpha \in \Delta^{+}} {\frac{H_\alpha} {(\alpha,\alpha)}}.
 \eqno(5.16)
$$
Note that this is analogous to the one in \c{DP}, where the standard
(spinless) elliptic Calogero-Moser system was considered.     
Clearly, 
 $\alpha (w) = (\alpha, \delta^\vee)$, where 
$$\delta^\vee = {1 \over 2} \sum_{\beta \in \Delta^+} \beta^\vee ,
\quad \beta^\vee = {2 \beta \over (\beta,\beta)}.\eqno(5.17)$$
If for $\alpha \in \Delta$, we write
$\alpha = \sum_{i=1}^{N} m^{i}_{\alpha} \alpha_{i}$,
then it is not hard to show that
$$l(\alpha) : = \alpha(w) = \sum_{i=1}^{N} m^{i}_{\alpha}.\eqno(5.18)$$ 
Therefore, $l(\alpha)$ is the level (or height) of $\alpha$.  Hence
$l(\alpha)$ is an integer, and assumes the value $1$ if and only if
$\alpha \in \pi$.

Now, with the definition of $x$ and $\eta_{\alpha}$ in (5.15), it
is easy to show that for $\alpha \in <\pi^{\prime}>^{+} $, we have
$$\frac {\xi_{\alpha} \xi_{-\alpha}} {sinh^{2} {1\over 2} \alpha(q)}
\sim 4\eta_{\alpha}\eta_{-\alpha}e^{-\alpha(x)-2\tau(l(\alpha)-1)}, \qquad
\tau \to \infty .\eqno(5.19)
$$
Therefore, if $\alpha \in <\pi^{\prime}>-\pi^{\prime}$, we have
$$\lim_{\tau \to \infty}{\frac{\xi_{\alpha} \xi_{-\alpha}} 
{sinh^{2} {1\over 2} \alpha(q)}} = 0. \eqno(5.20)
$$
On the other hand, if $\alpha \in \pi^{\prime}$, we obtain

$$\lim_{\tau \to \infty}{\frac{\xi_{\alpha} \xi_{-\alpha}} 
{sinh^{2} {1\over 2} \alpha(q)}} = 4 \eta_{\alpha} \eta_{-\alpha} 
e^{-\alpha(x)}.
\eqno(5.21)
$$
\smallskip
Accordingly, the scaling limit of the Hamiltonian
${\Cal H}$ of the
 hyperbolic spin Calogero-Moser system is given by
$$\eqalign{{\Cal H}^{s} (x,p,\eta) =& {1 \over 2} \sum_{i} p_{i}^{2} + 
          {1 \over 8} \sum_{i} \eta_{i}^{2} + {1 \over 2} \sum_{i}
           p_{i}\eta_{i} \cr
           & -\sum_{\alpha \in \pi^{\prime}} \eta_{\alpha}
           \eta_{-\alpha} e^ {-\alpha(x)}.\cr} \eqno(5.22)$$

Note that in contrast to the spinless case, we do not know a priori
the Poisson manifold on which ${\Cal H}^{s}$ is defined.  This
issue will be settled below, but first we shall work out the
scaling limits of the Hamiltonian equations of motion and the
(quasi) Lax equation in Proposition 5.4 which will in fact give
us some clue to this problem.

Let 
$$\eta = \sum_{i=1}^{N} \eta_{i} x_{i} + \sum_{\alpha \in \Delta}
  \eta_{\alpha} e_{\alpha}.\eqno(5.23)$$

\proclaim
{Proposition 5.7}  The scaling limit of the Hamiltonian equations
of motion in (5.7) is given by
$$\eqalign{& \dot x = p + {1 \over 2} \Pi_{\fh}\,\eta,\cr
           & \dot p = - \sum_{\alpha \in \pi^{\prime}}
             e^{-\alpha(x)} \eta_{\alpha} \eta_{-\alpha}H_{\alpha},\cr
           & \dot \eta = \bigl [\,\eta, {1\over 4} \Pi_{\fh}\,\eta
             + {1\over 2} p\,\bigr].\cr}\eqno(5.24)
$$
\endproclaim

\demo
{Proof}  The equation for $x$ is obvious from the equation for $q$.
On the other hand, the equation for $p$ is a consequence of our
previous analysis in (5.20)-(5.21) and the fact that
$\coth {1\over 2} \alpha (q) \to 1$  as $\tau \to \infty$
for $\alpha \in \Delta^{+}.$  To get the last equation above,
we make the substitution from (5.15) into the equation
for $\xi$ and then divide both sides by $e^{\tau}$, this gives
$$\aligned
         & (\Pi_{\fh^{\perp}}\,\eta)^{\cdot} \\
       = & \Bigl [\, e^{-\tau} \Pi_{\fh}\,\eta + \Pi_{\fh^{\perp}}\,\eta,
           {1\over 4} \Pi_{\fh}\,\eta + {1\over 2} p
          -{1\over 4} \sum_{\alpha \in <\pi^{\prime}>}
             \frac{\eta_{\alpha} e^{\tau}} {\sinh^{2} {1\over 2} 
            \alpha (x + 2 \tau w)}
             e_{\alpha}\,\Bigr]
\endaligned
$$
as $ (\Pi_{\fh}\,\eta)^{\cdot} = 0$.  Therefore, upon letting
$\tau \to \infty$, we find
$$(\Pi_{\fh^{\perp}}\,\eta)^{\cdot}
  = \bigl [\,\Pi_{\fh^{\perp}}\,\eta, {1\over 4} \Pi_{\fh}\,\eta + {1\over 2} p
     \,\bigr ].$$
Combining this with $(\Pi_{\fh}\,\eta)^{\cdot} = 0$, the equation for
$\eta$ follows.
\pf
\enddemo

At this juncture, we remark that the Lax operator $L(q,p,\xi)$ does
not actually admit a finite limit, as can be easily verified.  However, we
can remedy this by considering the following gauge-equivalent equation:
$$\eqalign{(Ad_{e^{-\tau w}} L)^{\cdot} =& [\, Ad_{e^{-\tau w}} L,
            Ad_{e^{-\tau w}} R(q) L \,]\cr
            & - Ad_{e^{-\tau w}} dR(q)(\Pi_{\fh}\,\xi) L.\cr}\eqno(5.25)$$
Thus we introduce
$$\eqalign{& L_{\tau} (x,p,\eta):=Ad_{e^{-\tau w}}
             L(x + 2 \tau w, p, \Pi_{\fh}\,\eta +
             e^{\tau} \Pi_{\fh^{\perp}}\,\eta),\cr
           & M_{\tau} (x,p,\eta):=Ad_{e^{-\tau w}} R(x+2\tau w)
             L(x + 2 \tau w, p, \Pi_{\fh}\,\eta +
             e^{\tau} \Pi_{\fh^{\perp}}\,\eta).\cr}\eqno(5.26)$$
Using the relation  
$Ad_{e^{-\tau w}} e_{\alpha} = e^{-\tau l(\alpha)} e_{\alpha}$ and
the $H$-equivariance of $R$, we easily find that
$$\eqalign{& L_{\tau}(x,p,\eta)\cr
         = & p + {1\over 2} \Pi_{\fh}\,\eta +
            \sum_{\alpha \in <\pi^{\prime}>} \frac{e^{{1\over 2}\alpha(x +
            2\tau w)}} {2 \sinh {1\over 2} \alpha (x + 2\tau w)}
            \eta_{\alpha} e^{-\tau(l(\alpha)-1)} e_{\alpha}\cr
            & + \sum_{\alpha \in {\overline \pi^{\prime}}^{+}}
              \eta_{\alpha} e^{-\tau(l(\alpha)-1)} e_{\alpha},\cr}\eqno(5.27)
$$
whereas
$$\eqalign{& M_{\tau}(x,p.\eta)\cr
         = & - {1\over 2} \sum_{\alpha \in <\pi^{\prime}>}
             \coth {1\over 2} \alpha (x + 2 \tau w)
             \frac{e^{{1\over 2}\alpha(x +
            2\tau w)}} {2 \sinh {1\over 2} \alpha (x + 2\tau w)}
            \eta_{\alpha} e^{-\tau(l(\alpha)-1)} e_{\alpha}\cr
           & + {1\over 2} \sum_{\alpha \in {\overline \pi^{\prime}}^{+}}
              \eta_{\alpha} e^{-\tau(l(\alpha)-1)} e_{\alpha}.\cr}\eqno(5.28)
$$
Now for $\alpha \in <\pi^{\prime}>^+$, we have
$$\frac {e^{{1\over 2}\alpha(x +
            2\tau w)}} {2 \sinh {1\over 2} \alpha (x + 2\tau w)}
             e^{-\tau(l(\alpha)-1)} \sim  e^{-\tau(l(\alpha)-1)},
          \quad   \tau \to \infty. \eqno(5.29)$$
Similarly, for  $\alpha \in <\pi^{\prime}>^-$, we obtain
$$\frac {e^{{1\over 2}\alpha(x +
            2\tau w)}} {2 \sinh {1\over 2} \alpha (x + 2\tau w)}
             e^{-\tau(l(\alpha)-1)} \sim - e^{\alpha(x)+\tau(l(\alpha)+1)},
          \quad   \tau \to \infty.\eqno(5.30)$$

\proclaim
{Proposition 5.8}  We have
$$\eqalign{\bL(x,p,\eta):=&\lim_{\tau \to \infty} L_{\tau}(x,p,\eta)\cr
  = & p + {1\over 2} \Pi_{\fh}\,\eta + \sum_{\alpha \in \pi} \eta_{\alpha}
      e_{\alpha} -\sum_{\alpha \in \pi^{\prime}} e^{-\alpha(x)}
      \eta_{-\alpha} e_{-\alpha},\cr}\eqno(5.31)$$
$$\eqalign{\bM(x,p,\eta):=&\lim_{\tau \to \infty} M_{\tau}(x,p,\eta)\cr
  = & -{1\over 2} \sum_{\alpha \in \pi} \eta_{\alpha}
      e_{\alpha} -{1\over 2} \sum_{\alpha \in \pi^{\prime}} e^{-\alpha(x)}
      \eta_{-\alpha} e_{-\alpha}.\cr}\eqno(5.32)$$
Moreover, the scaling limit of the (quasi) Lax equation (5.8) is given
by
$$\dot \bL = \bigl [\, \bL,\bM \,\bigr] = \bigl [\,\bL,{\bold R}(\bL)\,
  \bigr]\eqno(5.33)$$
where ${\bold R}$ is the constant $r$-matrix defined by
$${\bold R}(\eta) = {1\over 2}\sum_{\alpha \in \Delta^-} \eta_{\alpha}
 e_{\alpha} -{1\over 2} \sum_{\alpha \in \Delta^+} \eta_{\alpha}
 e_{\alpha}.\eqno(5.34)$$
\endproclaim

\demo
{Proof} Using the asymptotics in (5.29)-(5.30), we obtain

$$
\lim_{\tau \to \infty} \frac{e^{{1\over 2}\alpha(x +
            2\tau w)}} {2 \sinh {1\over 2} \alpha (x + 2\tau w)}
             e^{-\tau(l(\alpha)-1)} 
=\cases 0, &  \alpha \in <\pi^{\prime}>- (\pi^{\prime}\cup (-\pi^{\prime}))\\
 1, & \alpha \in \pi^{\prime} \\
 -e^{\alpha(x)}, & \alpha \in -\pi^{\prime}\endcases
$$
from which the formulas for $\bL$ and $\bM$ follow.  Therefore, in order
to demonstrate the validity of (5.33), it
remains to show that
 
$$\lim_{\tau \to \infty} 
Ad_{e^{-\tau w}} dR(x+2\tau w)(\Pi_{\fh}\,\eta)
 L(x + 2 \tau w, p, \Pi_{\fh}\,\eta +
             e^{\tau} \Pi_{\fh^{\perp}}\,\eta)\, =\, 0.$$
By the $H$-equivariance of $R$ and its explicit expression,
$$\aligned
         &Ad_{e^{-\tau w}} dR(x+2\tau w)(\Pi_{\fh}\,\eta)
 L(x + 2 \tau w, p, \Pi_{\fh}\,\eta +
             e^{\tau} \Pi_{\fh^{\perp}}\,\eta)\\
      =\, & dR(x+2\tau w)(\Pi_{\fh}\,\eta) L_{\tau}(x,p,\eta)\\
      =\, &  \sum_{\alpha \in <\pi^{\prime}>}
         \alpha(\Pi_{\fh}\,\eta) \eta_{\alpha}
        \frac {e^{{1\over 2}\alpha(x +
            2\tau w)}} {(2 \sinh {1\over 2} \alpha (x + 2\tau w))^3}
             e^{-\tau(l(\alpha)-1)} e_{\alpha}.
\endaligned
$$
But as $\tau \to \infty$, we have
$$\aligned
         &\frac {e^{{1\over 2}\alpha(x +
            2\tau w)}} {(2 \sinh {1\over 2} \alpha (x + 2\tau w))^3}
             e^{-\tau(l(\alpha)-1)}\\
         & \phantom{abc}\\
      \sim & \cases e^{-\alpha(x)-\tau(3l(\alpha)-1)},\,\, \alpha\in 
          <\pi^{\prime}>^+ \\
           -e^{2\alpha(x)+\tau(3l(\alpha)+1)}, \alpha \in <\pi^{\prime}>^-.
          \endcases
\endaligned
$$ 
Hence the required property follows.
\pf
\enddemo
\noindent{\bf Remark 5.9} (a) The constant $r$-matrix ${\bold R}$ 
is the scaling limit of the
dynamical $r$-matrix in the sense that
${\bold R}(\xi) = \lim_{\tau \to \infty} R(x+2\tau w) \xi.$
\smallskip
\noindent(b) It is a remarkable fact that the (quasi) Lax equation (5.8) 
scales to 
the genuine Lax equation in (5.33).  In other words, the obstruction
to integrability dissolves in the scaling limit.
\smallskip
\noindent(c) The reader should note that the scaling limit above is a singular
limit.  For this reason, the geometric structures are not preserved.
As the reader will see in what follows, ${\Cal H}^{s}$ is defined on
a Poisson manifold different from that of ${\Cal H}$.  Therefore, it
is not surprising that their Hamiltonian realization would require separate
consideration.
\smallskip
 We now describe a Hamiltonian formulation of the equations in
Proposition 5.7.  To do so, we consider the trivial Lie algebroid
$S = T\fh \times \fg$ over $\fh$, where $\fg$ is identified with
the semi-direct product $\fh \ltimes \fh^{\perp}$ associated
with the representation $ad$ of the Cartan subalgebra $\fh$
in $\fh^{\perp}$.  Thus the Lie algebroid bracket on $S$ is given
by
$$\eqalign{& [(Z,X),(Z',X')]_{S}(x) \cr
         = & (dZ'(x)Z(x)-dZ(x)Z'(x), dX'(x)Z(x)- dX(x) Z'(x)\cr
           & + [\,\Pi_{\fh} X(x), \Pi_{\fh^{\perp}} X'(x)\,]
             - [\,\Pi_{\fh} X'(x),\Pi_{\fh^{\perp}} X(x)\,])\cr}\eqno(5.35)$$
where  $Z,Z':\fh\longrightarrow \fh$ ,
$X,X': \fh\longrightarrow \fg$ are holomorphic maps and $x\in \fh$.

\proclaim
{Proposition 5.10} The Lie-Poisson structure on the dual bundle
$S^{*} \simeq \fh\times\fh\times \fg$ of the trivial Lie algebroid
$S$ is given by
$$\aligned
         &\{\,\varphi,\psi\,\}_{S^*} (x,p,\eta) \\
       = & (\delta_{1}\psi, \delta_{2}\varphi)-
           (\delta_{1}\varphi,\delta_{2}\psi) +
           (\eta, [\Pi_{\fh}\,\delta \varphi, \Pi_{\fh^{\perp}}\,\delta \psi])
          + [\Pi_{\fh^{\perp}}\, \delta\varphi, \Pi_{\fh}\,\delta \psi])
\endaligned
$$
and the Hamiltonian equations  generated by 
$\varphi :S^{*} \longrightarrow \Bbb C$ are:
$$\eqalign{
         & \dot x = \delta_{2} \varphi,\cr
         & \dot p = -\delta_{1}\varphi,\cr
         & \dot \eta = [\eta, \Pi_{\fh}\,\delta \varphi]
           + \Pi_{\fh}\, [\eta, \delta \varphi].\cr}\eqno(5.36)$$
\endproclaim

\demo
{Proof} Using the method of calculation in Section 2, we have
$$\eqalign {& \{\,\varphi, \psi\,\}_{S^*} (x,p,\eta)\cr
          = & l_{[s(\varphi),s(\psi)]_{S}} (x,p,\eta)
              +(\delta_{1}\psi, \delta_{2}\varphi)-
           (\delta_{1}\varphi,\delta_{2}\psi).\cr}$$
 Now, from the expression for $[\cdot,\cdot]_{S}$,
it is easy to check that
$$\aligned
         &[s(\varphi),s(\psi)]_{S}(x)\\
       = &(0,  [\Pi_{\fh}\delta\varphi,\Pi_{\fh^{\perp}}\delta\psi]
          -  [\Pi_{\fh} \delta \psi, \Pi_{\fh^{\perp}}\delta\varphi]).
\endaligned
$$
Hence we have 
$$\aligned
         & l_{[s(\varphi),s(\psi)]_{S}}(x,p,\eta)\\
       = & (\eta,  [\Pi_{\fh}\,\delta\varphi,\Pi_{\fh^{\perp}}\,\delta\psi]
          + [\Pi_{\fh^{\perp}}\, \delta\varphi, \Pi_{\fh}\,\delta\psi]).
\endaligned
$$ 
Assembling the calculations, we obtain the formula for
$\{\,\varphi,\psi\,\}_{S^*} (x,p,\eta)$.  
\pf
\enddemo

To prepare for our next result, we need to introduce further constructs.
First of all, let ${\bold A}^{*} \simeq
T\fh \times \fg$ be the coboundary dynamical Lie algebroid
associated with the constant r-matrix ${\bold R}$.  Since $\fh$
is Abelian, the Lie-Poisson structure on its dual bundle ${\bold A}
\simeq T\fh \times \fg$ takes the form
$$\eqalign{& \{\varphi, \psi\}_{{\bold A}} (x,p,\eta)\cr
        =  & (\eta, [{\bold R} (\delta \varphi) - \delta_{2}\varphi,
              \delta \psi] + [\delta \varphi, {\bold R} (\delta \psi)
              - \delta_{2} \psi])\cr
           & + (\delta_{1}\psi, \Pi_{\fh}\delta\varphi)-
           (\delta_{1}\varphi,\Pi_{\fh}\delta\psi).\cr}\eqno(5.37)
$$
Now, define
$$\eqalign{&{\boldsymbol \rho}: T\fh\times \fg \simeq S^{*}\longrightarrow 
           {\bold A}\simeq T\fh \times \fg\cr
           & (x,p,\eta)\mapsto (x,-\Pi_{\fh} \eta, \bL(x,p,\eta)).\cr}
           \eqno(5.38)$$

\proclaim
{Theorem 5.11}  ${\boldsymbol \rho}$ is an $H$-equivariant Poisson map,
where the $H$-action on $S^*$ given by
$$
\aligned
      h\cdot(x,p,\eta)& = (x,p, Ad_{h} \eta)\\
                  & = (x,p, \Pi_{\fh}\,\eta + \Pi_{\fh^{\perp}} Ad_{h} \eta)
\endaligned
$$
is Hamiltonian with equivariant momentum map 
${\bold J}:T\fh \times \fg \longrightarrow \fh$, 
$(x,p,\eta) \mapsto -\Pi_{\fh}\,\eta$.  Moreover, the equations in
Proposition 5.7 are the Hamiltonian equations generated by
${\Cal H}^{s} (x,p,\eta) = \bL^{*} Q(x,p,\eta)$ in
the Lie-Poisson structure $\{\,\cdot,\cdot\,\}_{S^{*}}$ and
admit $\bL^{*} I(\fg)$ as a family of conserved quantities
in involution.
\endproclaim 
            
\demo
{Proof}  Let $\varphi$, $\psi \in C^{\infty} ({\bold A})$.  By direct
calculation, we have
$$
\aligned
       &\delta(\varphi\circ\rho)(x,p,\eta)\\
     = &-\delta_{2}\varphi(\rho(x,p,\eta)) + {1\over 2} \Pi_{\fh}\,
         \delta \varphi (\rho(x,p,\eta)) \\
       &+ \sum_{\alpha \in \pi} (\delta \varphi(\rho(x,p,\eta)))_{-\alpha}
         e_{-\alpha} -
        \sum_{\alpha \in \pi^{\prime}} e^{-\alpha(x)} (\delta\varphi
        (\rho(x,p,\eta))_{\alpha} e_{\alpha},
\endaligned
$$
$\delta_{1}(\varphi\circ\rho)(x,p,\eta)
      = \delta_{1} \varphi (\rho(x,p,\eta) + \sum_{\alpha \in \pi^{\prime}} 
         {\delta \varphi}_{\alpha} \eta_{\alpha} e^{-\alpha(x)} H_{\alpha},$
and 
$\delta_{2} (\varphi\circ\rho)(x,p,\eta) = 
\Pi_{\fh}\,\delta\varphi(\rho(x,p,\eta)).$
To simplify notation, let $X = \delta \varphi(\rho(x,p,\eta))$,
$Y = \delta_{1} \varphi (\rho(x,p,\eta))$ and $Z=\delta_{2}\varphi
(\rho(x,p,\eta))$ and denote the corresponding quantities associated
with $\psi$ by $X'$, $Y'$ and $Z'$ respectively. Then it follows
from the expression of $\{\,\cdot,\cdot\,\}_{S^{*}}$ in Proposition 5.10
and the above calculation that
$$
\aligned
       & \{\varphi\circ\rho, \psi\circ\rho\}_{S^{*}} (x,p,\eta)\\
    =  & \sum_{\alpha \in \pi^{\prime}} e^{-\alpha(x)} \eta_{-\alpha}
         X'_{\alpha} \alpha \bigl(Z+{1\over 2} \Pi_{\fh}\,X\bigr)
        + \sum_{\alpha \in \pi} \eta_{\alpha} X'_{-\alpha}\, 
        \alpha \bigl(Z - {1\over 2} \Pi_{\fh}\,X\bigr)\\
       & +  (Y', \Pi_{\fh}\,X) - (X\leftrightarrow X', Y\leftrightarrow Y',
         Z\leftrightarrow Z').
\endaligned
$$
On the other hand,
$$
\aligned
       &\{\varphi, \psi\}_{{\bold A}}\circ \rho (x,p,\eta)\\
     = & (\bL(x,p,\eta), [{\bold R}(X) -Z,X'] + [X,{\bold R}(X')-Z'])\\
       & + (Y', \Pi_{\fh}\,X) - (Y, \Pi_{\fh}\, X').
\endaligned
$$
But
$$
\aligned
       & \Bigl(\sum_{\alpha\in \pi} \eta_{\alpha} e_{\alpha},
           [{\bold R}(X) -Z,X'] + [X,{\bold R}(X')-Z']\Bigr)\\
     = & \sum_{\alpha\in \pi} \eta_{\alpha} X'_{-\alpha}
         \alpha \bigl(Z - {1\over2} \Pi_{\fh}\,X \bigr) -
         (X\leftrightarrow X', Y\leftrightarrow Y',
         Z\leftrightarrow Z'),
\endaligned
$$
while
$$
\aligned
       & \Bigl(-\sum_{\alpha\in \pi^{\prime}} e^{-\alpha(x)} \eta_{\alpha} 
          e_{\alpha}, [{\bold R}(X) -Z,X'] + [X,{\bold R}(X')-Z']\Bigr)\\
     = & \sum_{\alpha\in \pi^{\prime}} e^{-\alpha(x)} \eta_{-\alpha}
         X'_{\alpha} \alpha \bigl(Z + {1\over 2} \Pi_{\fh}\,X \bigr)
         -(X\leftrightarrow X', Y\leftrightarrow Y',
         Z\leftrightarrow Z').
\endaligned
$$
Putting the calculations together, we conclude that
$\{\varphi, \psi\}_{{\bold A}}\circ \rho (x,p,\eta)$ is identical
to $\{\varphi\circ\rho, \psi\circ\rho\}_{S^{*}} (x,p,\eta)$.
Thus $\rho$ is a Poisson map.  Alternatively, we can also establish
the assertion by showing that
the dual of the bundle map ${\boldsymbol \rho}$ is a morphism of Lie
algebroids.  

To show that the equations in Proposition 5.7 are the Hamiltonian
equations generated by ${\Cal H}^{s}$ in the Poisson structure
$\{\,\cdot,\cdot \,\}_{S^{*}}$, note that  
$$\delta_{1} {\Cal H}^{s} = \sum_{\alpha \in \pi^{\prime}}
\eta_{\alpha} \eta_{-\alpha} e^{-\alpha(x)} H_{\alpha},\quad
\delta_{2} {\Cal H}^{s} = p + {1\over 2} \Pi_{\fh}\,\eta$$
and 
$$\delta {\Cal H}^{s} = {1\over 2} p + {1\over 4} \Pi_{\fh}\,\eta
  - \sum_{\alpha \in \pi^{\prime}} e^{-\alpha(x)}(\eta_{\alpha}
  e_{\alpha} + \eta_{-\alpha} e_{-\alpha}).$$
From these formulas, it is clear that the equations for $x$ and
$p$ from (5.36) with $\varphi = {\Cal H}^{s}$ are identical to
the corresponding ones in Proposition 5.7.  To show that the equations for 
$\eta$
are identical as well, it is enough to check that
$[\,\eta, \delta {\Cal H}^{s}\,] \in \fh^{\perp}$.
Write
$\eta = \Pi_{\fh}\,\eta + \Pi_{\fh^{\perp}}\,\eta$ and substitute
into $[\,\eta, \delta {\Cal H}^{s}\,]$.  As
$[\,\fh, \fh^{\perp}\,] \subset \fh^{\perp}$, we only have to consider
the term
$[\, \Pi_{\fh^{\perp}}\,\eta, -\sum_{\alpha \in \pi^{\prime}} 
 e^{-\alpha(x)} (\eta_{\alpha}
  e_{\alpha} + \eta_{-\alpha} e_{-\alpha})\,]$ in
$[\,\eta, \delta {\Cal H}^{s}\,]$.  Expanding out, we have
$$
\aligned
       &[\,\Pi_{\fh^{\perp}}\,\eta, -\sum_{\alpha \in \pi^{\prime}} 
            e^{-\alpha(x)} (\eta_{\alpha}
            e_{\alpha} + \eta_{-\alpha} e_{-\alpha})\,]\\
     =  &-\sum_{\alpha\in \pi^{\prime}}\sum_{\beta\in \Delta^{+}}
           e^{-\alpha(x)} \eta_{\alpha}\eta_{\beta}[e_{\beta},
           e_{\alpha}] - \sum_{\alpha\in \pi^{\prime}}\sum_{\beta\in 
           \Delta^{+}}
           e^{-\alpha(x)} \eta_{-\alpha}\eta_{-\beta}[e_{-\beta},
           e_{-\alpha}] \\
          & - \sum_{\alpha\in \pi^{\prime}}\sum_{\beta\in \Delta^{+}}
           e^{-\alpha(x)} \eta_{-\alpha}\eta_{\beta}[e_{\beta},
           e_{-\alpha}] - \sum_{\alpha\in \pi^{\prime}}\sum_{\beta\in 
           \Delta^{+}}
           e^{-\alpha(x)} \eta_{\alpha}\eta_{-\beta}[e_{-\beta},
           e_{\alpha}]. 
\endaligned
$$
Clearly, the first two terms of the above sum are in $\fh^{\perp}$.
Now, 
$$\aligned
         &\Pi_{\fh}\Bigl( \sum_{\alpha\in \pi^{\prime}}\sum_{\beta\in \Delta^+}
           e^{-\alpha(x)} \eta_{-\alpha}\eta_{\beta}[e_{\beta},
           e_{-\alpha}]\Bigr)\\
       = & \sum_{\alpha \in \pi^{\prime}}
           e^{-\alpha(x)} \eta_{\alpha}\eta_{-\alpha} H_{\alpha}
\endaligned
$$ 
while
$$\aligned
         &\Pi_{\fh}\,\Bigl(\sum_{\alpha\in \pi^{\prime}}
          \sum_{\beta\in \Delta^+}
           e^{-\alpha(x)} \eta_{\alpha}\eta_{-\beta}[e_{-\beta},
           e_{\alpha}]\Bigr)\\
       = &- \sum_{\alpha \in \pi^{\prime}}
           e^{-\alpha(x)} \eta_{\alpha}\eta_{-\alpha} H_{\alpha}.
\endaligned
$$
So the sum of the last two terms in the above sum is in 
$\fh^{\perp}$ as well.   We shall leave the rest of the proof to
the reader.
\pf
\enddemo

We shall call the Hamiltonian systems generated by ${\Cal H}^{s}$
in the Lie-Poisson structure $\{\,\cdot,\cdot\,\}_{S^*}$ {\it spin
Toda lattices}.
To close this section, we shall consider reduction of the spin Toda lattices
.
As before, we consider the submanifold ${\Cal U}$ defined in (5.9).  Then
clearly, the $H$-action defined in Theorem 5.11 induces a Hamiltonian
action on $T\fh \times {\Cal U}$.  Denote the corresponding momentum map
also by ${\bold J}$, we have ${\bold J}^{-1}(0) = T\fh\times (\fh^{\perp}\cap
{\Cal U})$.  In this case, it is easy to verify that a generic orbit
${\Cal O}\subset \fg$ (recall that $\fg \simeq \fh \ltimes \fh^{\perp}$)
is of dimension $2N$, where $N = rank (\fg)$.  Therefore,
${\Cal O}_{red} = ({\Cal O} \cap {\Cal U} \cap \fh^{\perp})/H$ is a point.
Indeed, we have

\proclaim
{Corollary 5.13} The reduction of the Hamiltonian ${\Cal H}^{s}$
of the spin Toda lattice on $T\fh \times {\Cal O}_{red}$ is given by
$${\Cal H}^{s}_0 (x,p) =  {1 \over 2} \sum_{i} p_{i}^{2}  
         -\sum_{\alpha \in \pi^{\prime}} c_{\alpha}
          e^ {-\alpha(x)} \eqno(5.39)$$
where $c_{\alpha} = s_{-\alpha}$ is a constant.  Thus the Hamiltonian 
equations of motion generated
by ${\Cal H}^{s}_0$ are:
$$\eqalign{
       & \dot x = p, \cr
       & \dot p = - \sum_{\alpha \in \pi^{\prime}} c_{\alpha}
         e^{-\alpha(x)} H_{\alpha}.\cr}\eqno(5.40)$$
\endproclaim
\pf

Hence by reduction, we obtain a family of Toda lattices parametrized by 
subsets $\pi^{\prime}$ of $\pi$. 

\bigskip
\bigskip

\subhead
6. \ Solution of the hyperbolic spin Calogero-Moser systems and the
\linebreak \phantom{abc}\,\,\, spin Toda lattices
\endsubhead

\subsubhead (a) The hyperbolic spin Calogero-Moser systems
\endsubsubhead

\bigskip
We begin by solving the equation
$$\eqalign{
  &{d\over dt} \, (q, 0, L(q,p,\xi))\cr
 =\,&(p, 0, [\,L(q,p,\xi), R(q) L(q,p,\xi)\,]).\cr}\eqno(6.1)
$$
for the hyperbolic spin Calogero-Moser system which we obtain
from Proposition 5.4 by restricting to the invariant manifold
$J^{-1} (0)$.  As the reader will see, this will lead us eventually
to the solution of the associated integrable model, whose equations
are given in Proposition 5.6.

In order to set up the factorization problem properly, it is
necessary to have precise knowledge of the Lie algebroids and
Lie groupoids involved.  Let us begin to describe these objects.  Let
${\frak b}^{-} =\fh + \sum_{\alpha \in \Delta^{-}} \fg_{\alpha}$
and ${\frak b}^{+} =\fh + \sum_{\alpha \in \Delta^{+}} \fg_{\alpha}$
be opposing Borel subalgebras of $\fg$.  From the definition of $R$ in
(5.1), we have 
$$\eqalign{R^{\pm}(q)\xi = &\pm {1\over 2}\sum_{i} \xi_{i} x_{i}-
 {1\over 2}\sum_{\alpha \in <\pi^{\prime}>} {\frac{e^{\mp{1\over 2}\alpha(q)}}
 {\sinh {1\over 2}\alpha (q)}} \xi_{\alpha} e_{\alpha}\cr
& \pm \sum_{\alpha \in {\overline \pi^{{\prime}^{\mp}}}} \xi_{\alpha} e_{\alpha}.
\cr}\eqno(6.2)$$
Therefore, $R^{+}(q)\xi$ is in the parabolic subalgebra
$${\frak p}^{-}_{\pi^{\prime}} = {\frak b}^{-} + \sum_{\alpha\in 
<\pi^{\prime}>^+}\, \fg_{\alpha} \eqno(6.3)$$
containing ${\frak b}^-$, while $R^{-}(q)\xi$ is in the parabolic
subalgebra
$${\frak p}^{+}_{\pi^{\prime}} = {\frak b}^{+} + \sum_{\alpha\in 
<\pi^{\prime}>^-}\, \fg_{\alpha} \eqno(6.4)$$
containing ${\frak b}^+$.  Now, recall that we have the standard
decompositions \c{Kn}
$${\frak p}^{\pm}_{\pi^{\prime}} = \fg_{\pi^{\prime}} + 
{\frak n}^{\pm}_{\pi^{\prime}}\eqno(6.5)$$
where 
$$\fg_{\pi^{\prime}} = \fh + \sum_{\alpha \in <\pi^{\prime}>} {\fg_{\alpha}}
\eqno(6.6)$$
is the Levi factor of ${\frak p}^{\pm}_{\pi^{\prime}}$, and
$${\frak n}^{\pm}_{\pi^{\prime}} = \sum_{\alpha \in
{\overline \pi^{{\prime}^{\pm}}}} \fg_{\alpha} \eqno(6.7)$$
are the nilpotent radicals.  Moreover, we have the identity
$$\fg = {\frak n}^{-}_{\pi^{\prime}} + {\fg}_{\pi^{\prime}} +
{\frak n}^{+}_{\pi^{\prime}}.\eqno(6.8)$$

Let $P^{\pm}_{\pi^{\prime}}$, $G_{\pi^{\prime}}$ and $N^{\pm}_{\pi^{\prime}}$
be the simply-connected Lie subgroups of $G$ with corresponding Lie
subalgebras  ${\frak p}^{\pm}_{\pi^{\prime}}$, $\fg_{\pi^{\prime}}$
and ${\frak n}^{\pm}_{\pi^{\prime}}$.  Then we have
$$P^{\pm}_{\pi^{\prime}} = N^{\pm}_{\pi^{\prime}} G_{\pi^{\prime}}\eqno(6.9)$$
and the submanifold
$$G_{0}={N^{-}_{\pi^{\prime}}} {G_{\pi^{\prime}}} {N^{+}_{\pi^{\prime}}}
\eqno(6.10)$$
is an open dense subset of $G$.

From the explicit expression for $R^+$ in (6.2) and the definition
of ${\Cal R}^+$, we find that
$$Im {\Cal R}^{+} = \bigcup_{q\in U} \lbrace 0_{q} \rbrace \times
{\frak p}^{-}_{\pi^{\prime}}\times \fh . \eqno(6.11)$$
Similarly, we have 
$$Im {\Cal R}^{-} = \bigcup_{q\in U} \lbrace 0_{q} \rbrace \times
{\frak p}^{+}_{\pi^{\prime}}\times \fh . \eqno(6.12)$$
Therefore, the unique source-simply connected Lie groupoids of
$Im {\Cal R}^{\pm}$ are given by
$$ \Gamma_{\pm} = U \times P^{\mp}_{\pi^{\prime}}\times U.\eqno(6.13)$$

We next describe the ideals ${\Cal I}^{\pm}$ of (4.7) for the case
under consideration.

\proclaim
{Lemma 6.1}  ${\Cal I}^{\pm} = \bigcup_{q\in U} \lbrace 0_q \rbrace
\times {\frak n}^{\mp}_{\pi^{\prime}} \times \lbrace 0 \rbrace.$
\endproclaim

\demo
{Proof} We shall give the proof for ${\Cal I}^{-}$.  Suppose
$(0_q,X,0)\in {\Cal I}^{-}$, then there exists $Z\in \fh$ such
that ${\Cal R}^{+}(0_q,X,Z)=0$.  Equivalently, we have
$-\iota Z + R^{+}(q) X =0$ and $\Pi_{\fh} X = 0$.  But from the
explicit expression for $R^+$ in (6.1), we easily find that
$X_{\alpha}=0$ for $\alpha \in \Delta^{-}\,\cup <\pi^{\prime}>^+$.
This shows that $X\in {\frak n}^{+}_{\pi^{\prime}}$.  The converse
is clear by reversing the steps in the above argument.
\pf
\enddemo

From this lemma, it follows that
$$\eqalign{Im {\Cal R}^{+}/{\Cal I}^{+}& = \bigcup_{q\in U} 
  \lbrace 0_q \rbrace \times ({\frak p}^{-}_{\pi^{\prime}}/
  {\frak n}^{-}_{\pi^{\prime}}) \times \fh\cr
  & \simeq \bigcup_{q\in U} \lbrace 0_q \rbrace \times
  \fg_{\pi^{\prime}} \times \fh\cr}\eqno(6.14)
$$
where the identification map is given by
$$(0_q,X+{\frak n}^{-}_{\pi^{\prime}},Z)\mapsto
  (0_q, \Pi^{-}_{\fg_{\pi^{\prime}}}X,Z).\eqno(6.15)$$
Here, $\Pi^{-}_{\fg_{\pi^{\prime}}}$ is the projection onto
$\fg_{\pi^{\prime}}$ relative to the direct sum decomposition
${\frak p}^{-}_{\pi^{\prime}} = \fg_{\pi^{\prime}} + 
{\frak n}^{-}_{\pi^{\prime}}$.  
Similarly,
$$\eqalign{Im {\Cal R}^{-}/{\Cal I}^{-}& = \bigcup_{q\in U} 
  \lbrace 0_q \rbrace \times ({\frak p}^{+}_{\pi^{\prime}}/
  {\frak n}^{+}_{\pi^{\prime}}) \times \fh\cr
  & \simeq \bigcup_{q\in U} \lbrace 0_q \rbrace \times
  \fg_{\pi^{\prime}} \times \fh.\cr}\eqno(6.16)
$$
This time, the identification is given by the map
$$(0_q,X+{\frak n}^{+}_{\pi^{\prime}},Z)\mapsto
  (0_q, \Pi^{+}_{\fg_{\pi^{\prime}}}X,Z).\eqno(6.17)$$
and $\Pi^{+}_{\fg_{\pi^{\prime}}}$ is the projection onto
$\fg_{\pi^{\prime}}$ relative to 
${\frak p}^{+}_{\pi^{\prime}} = \fg_{\pi^{\prime}} + 
{\frak n}^{+}_{\pi^{\prime}}.$

\proclaim
{Proposition 6.2}  The isomorphism $\theta : Im {\Cal R}^{+}/{\Cal I}^{+}
\longrightarrow Im {\Cal R}^{-}/{\Cal I}^{-}$ defined in Proposition 4.4
is given by
$$\theta (0_q,\Pi^{-}_{\fg_{\pi^{\prime}}} X, Z) =
    (0_q, -\iota Z + Ad_{e^q}\Pi^{-}_{\fg_{\pi^{\prime}}} X, Z)
    \eqno(6.18)$$
for all $q\in U$, $X\in {\frak p}^{-}_{\pi^{\prime}}$ and
$Z\in \fh.$
\endproclaim

\demo
{Proof} From the expression for $R^{\pm}(q)\xi$, we have

$$\aligned
        & \theta (0_q,-\iota Z' +{1\over 2}\Pi_{\fh} \xi
          -{1\over 2}\sum_{\alpha \in <\pi^{\prime}>} 
          {\frac{e^{-{1\over 2}\alpha(q)}}
          {\sinh {1\over 2}\alpha (q)}} \xi_{\alpha} e_{\alpha},
          \Pi_{\fh} \xi) \\
      =& (0_q,-\iota Z' -{1\over 2}\Pi_{\fh} \xi
          -{1\over 2}\sum_{\alpha \in <\pi^{\prime}>} 
          {\frac{e^{{1\over 2}\alpha(q)}}
          {\sinh {1\over 2}\alpha (q)}} \xi_{\alpha} e_{\alpha},
          \Pi_{\fh} \xi).
\endaligned
$$
The assertion then follows from the identity 
$Ad_{e^q} e_{\alpha} = e^{\alpha (q)} e_{\alpha}.$
\pf
\enddemo
We shall make the natural identifications
$N^{\pm}_{\pi^{\prime}}\backslash P^{\pm}_{\pi^{\prime}}
\simeq G_{\pi^{\prime}}$ using the relation in (6.9) in what follows.

\proclaim
{Corollary 6.3}  The isomorphism $\theta$ can be lifted up to a
Lie groupoid isomorphism
$$\eqalign{\Theta :& U\times (N^{-}_{\pi^{\prime}}\backslash 
P^{-}_{\pi^{\prime}})
\times U
\longrightarrow U\times (N^{+}_{\pi^{\prime}}\backslash P^{+}_{\pi^{\prime}}) 
\times U\cr
& (u,\bl^{-}(g),v)\mapsto (u, e^{u}\bl^{-}(g)e^{-v},v)\cr}
\eqno(6.19)$$
where for $g\in P^{-}_{\pi^{\prime}}$, $\bl^{-}(g)\in G_{\pi^{\prime}}$
denotes the factor in the unique factorization 
$g = \bn^{-}(g) \bl^{-}(g)$, $\bn^{-}(g)\in N^{-}_{\pi^{\prime}}.$
\endproclaim

\demo
{Proof} This is straightforward verification.
\pf
\enddemo

We are now ready to solve Eqn.(6.1).  To do so, we have to solve
the factorization problem 
$$exp \{\, t(0,0,L(q_0,p_0,\xi_0))\}(q_0)
      =\,\gamma_{+} (t)\,\gamma_{-} (t)^{-1}\eqno(6.20)$$
for 
$(\gamma_{+}(t),\gamma_{-}(t))=
((q_0, k_{+} (t), q(t)), (q_0, k_{-} (t), q(t))) \in
Im ({\Cal R}^{+}, {\Cal R}^{-})$
satisfying
the condition in (4.19), where
$(q_0,p_0,\xi_0) \in J^{-1}(0) = TU \times ({\Cal U} \cap \fh^{\perp})$
is the initial value of $(q,p,\xi)$.
 We shall use the following notation (analogous
to what we did in Corollary 6.3): for $g\in  P^{+}_{\pi^{\prime}}$,
$\bn^{+}(g)\in N^{+}_{\pi^{\prime}}$, $\bl^{+}(g)\in G_{\pi^{\prime}}$
will denote the factors in the unique factorization
$g = \bn^{+}(g) \bl^{+}(g)$.

With the notation above, we have $k_{\pm}(t) \in  P^{\mp}_{\pi^{\prime}}.$
Therefore, the relation
$e^{t L(q_0,p_0,\xi_0)} = k_{+}(t)k_{-}(t)^{-1}$ (which follows from
(6.20)) can be rewritten 
in the form
$$e^{t L(q_0,p_0,\xi_0)}=\bn^{-}(k_{+}(t))\bl^{-}(k_{+}(t))\bl^{+}(k_{-}(t))^{-1}
  \bn^{+}(k_{-}(t))^{-1}.$$
But from Theorem 4.5 (b) and Corollary 6.3, we have 
$\bl^{+}(k_{-}(t)) = e^{q_{0}}\bl^{-}(k_{+}(t))e^{-q(t)}.$  Hence it
follows that
$$\eqalign{& e^{tL(q_0,p_0,\xi_0)}\cr
        = & \bn^{-}(k_{+}(t))(\bl^{-}(k_{+}(t))
           e^{q(t)}\bl^{-}(k_{+}(t))^{-1}e^{-q_{0}})
           \bn^{+}(k_{-}(t))^{-1}.\cr}\eqno(6.21)$$
where the middle factor is in $G_{\pi^{\prime}}$ and
$\bn^{\mp}(k_{\pm}(t)) \in N^{\mp}_{\pi^{\prime}}.$
We shall obtain the factors $\bn^{\mp}(k_{\pm}(t))$, $\bl^{-}(k_{+}(t))$
and $q(t)$ in several steps.  First of all, from the fact that 
$e^{t L(q_0,p_0,\xi_0)}\in P^{+}_{\pi^{\prime}}$, we can find
 (as a consequence of (6.9)) unique  $g(t) \in  G_{\pi^{\prime}}$,
$n_{+}(t) \in N^{+}_{\pi^{\prime}}$ satisfying $n_{+}(0) =g(0) =1$
such that
$$e^{t L(q_0,p_0,\xi_0)} = g(t)n_{+}(t)^{-1}.\eqno(6.22)$$
By comparing (6.21) and (6.22), we obtain
$$\bn^{-}(k_{+}(t)) = 1,\quad \bn^{+}(k_{-}(t)) = n_{+}(t).\eqno(6.23)$$
Hence (6.21) reduces to the factorization problem 
$$g(t) e^{q_{0}} = \bl^{-}(k_{+}(t))
           e^{q(t)}\bl^{-}(k_{+}(t))^{-1}.\eqno(6.24)$$  
Since $G_{\pi^{\prime}}$ is a reductive Lie group, we can find
(at least for small values of $t$)
$x(t)\in G_{\pi^{\prime}}$ (unique up to transformations 
$x(t)\rightarrow x(t) \delta (t)$, where $\delta (t) \in H$) and
unique $d(t)\in H$ such that
$$g(t) e^{q_{0}} = x(t) d(t) x(t)^{-1}\eqno(6.25)$$
with $x(0)=1$, $d(0)=e^{q_{0}}$.
This uniquely determines $q(t)$ via the formula
$$q(t) = log \,d(t).\eqno(6.26)$$
On the other hand, let us fix one such $x(t)$.  We shall seek
$\bl^{-}(k_{+}(t))$ in the form
$$\bl^{-}(k_{+}(t)) = x(t) b(t), \quad b(t)\in H.\eqno(6.27)$$
To determine $b(t)$, we impose the condition in (4.19 a).
After some calculations, we find that $b(t)$ satisfies the equation
$$\dot b(t) = T_{e} l_{b(t)}\bigl ({1\over 2} \dot q(t) - \Pi_{\fh} 
(T_{x(t)} l_{{x(t)}^{-1}} \dot x(t)) \bigr )\eqno(6.28)$$
with $b(0)=1$.  Solving the equation explicitly, we find that
$$\bl^{-}(k_{+}(t)) = x(t) exp\{{1\over 2}(q(t) - q_{0})-
\int _{0}^{t}{\Pi_{\fh} 
(T_{x(\tau)} l_{{x(\tau)}^{-1}} 
\dot x(\tau))}\, d\tau \} .\eqno(6.29)$$
Combining (6.23) and (6.29), we finally have
$$k_{+}(t) =  x(t) exp\{{1\over 2}(q(t) - q_{0})-
\int _{0}^{t}{\Pi_{\fh} 
(T_{x(\tau)} l_{{x(\tau)}^{-1}} 
\dot x(\tau))}\, d\tau \} .\eqno(6.30)$$
Hence we can write down the solution of Eqn.(6.1) by using (4.20).
Note, however, we cannot determine  $\xi(t)$ from
the solution for $L$ as the expression for $L$ does not involve
$\xi_{\alpha}$ for $\alpha \in {\overline \pi^{\prime}}^-$. The 
solution of Eqn.(5.7) on $J^{-1}(0)$ is given in the following.

\proclaim
{Theorem 6.4}  Let $(q_0,p_0,\xi_0)\in J^{-1}(0) = TU \times ({\Cal U} \cap
\fh^{\perp})$.  Then the Hamiltonian flow on $J^{-1}(0)$ generated by
$$\eqalign{{\Cal H} (q,p,\xi) =& {1 \over 2} \sum_{i} p_{i}^{2} +
         {1 \over 8} \sum_{i} \xi_{i}^{2} + {1 \over 2} \sum_{i}
         p_{i}\xi_{i} \cr
         &-{1 \over 8} \sum_{\alpha \in <\pi^{\prime}>} \frac{\xi_{\alpha}
         \xi_{-\alpha}} {sinh^{2} {1 \over 2} \alpha (q)}\cr}
$$ 
with initial condition $(q(0),p(0),\xi(0)) = (q_0,p_0,\xi_0)$ is given
by
$$\eqalign{& q(t) =\, log\,d(t), \cr
       & \xi (t) =\, Ad_{k_{+}(t)^{-1}} \xi_{0},\cr
       & p(t) =\, Ad_{k_{\pm}(t)^{-1}} L(q_0,p_0,\xi_0)
         -{1\over 2} \sum_{\alpha \in <\pi^{\prime}>}
         \frac{e^{{1\over 2} \alpha (q(t))}}
         {\sinh {1\over 2} \alpha (q(t))} \xi_{\alpha} (t) e_{\alpha}\cr
       &\,\,\,\,\,\,\,\,\,\, - \sum_{\alpha \in {\overline \pi^{\prime}}^{+}}
         \xi_{\alpha} (t) e_{\alpha}\cr}\eqno(6.31)
$$
where $d(t)$, $k_{\pm} (t)$ are constructed from the above procedure
and in the expression for $p(t)$, the quantities $q(t)$ and $\xi(t)$ 
which appear on the right hand side are given
by the formulas above that expression.
\endproclaim

\demo
{Proof} We first show $ \xi (t) =\, Ad_{k_{+}(t)^{-1}} \xi_{0}$ solves
the equation $\dot \xi = [\, \xi, R^{+}(q) L(q,p,\xi)\,]$ in
Proposition 5.4.  To do so, we differentiate the expression
for $\xi (t)$, this gives
$$\dot \xi (t) = \bigl [T_{{k_{+}(t)}^{-1}} r_{k_{+}(t)} {d\over dt} k_{+}(t)^{-1},
  \xi (t) \bigr].$$
But
$$
\aligned
       &T_{{k_{+}(t)}^{-1}} r_{k_{+}(t)} {d\over dt} k_{+}(t)^{-1}\\
   = \,& -T_{k_{+}(t)} l_{{k_{+}(t)}^{-1}} \dot k_{+} (t)\\
   = \,&- R^{+}(q(t)) L(q(t), p(t), \xi(t))
\endaligned
$$
from the argument in Theorem 4.7.  Hence the claim.  To get the
formula for $p(t)$, we simply equate the following two expressions for
$L((q(t),p(t),\xi(t))$, namely,
$L(q(t),p(t),\xi(t)) = Ad_{{k_{\pm}(t)}^{-1}} L(q_0,p_0,\xi_0)$ and
$$
\aligned
        L(q(t),p(t),\xi(t)) =&\, p(t) + {1\over 2} 
        \sum_{\alpha \in <\pi^{\prime}>}
         \frac{e^{{1\over 2} \alpha (q(t))}}
         {\sinh {1\over 2} \alpha (q(t))} \xi_{\alpha} (t) e_{\alpha}\\
        & + \sum_{\alpha \in {\overline \pi^{\prime}}^{+}}
         \xi_{\alpha} (t) e_{\alpha}.
\endaligned
$$
This completes the proof.
\pf
\enddemo

We next turn to the solution of the associated integrable model
on $TU \times \fg_{red}$ with Hamiltonian
${\Cal H}_{0} (q,p,s) = {1 \over 2} \sum_{i} p_{i}^{2} -{1\over 4}
\sum_{\alpha \in <\pi^{\prime}>^{+}} \frac {s_{\alpha}
         s_{-\alpha}} {sinh^{2} {1 \over 2} \alpha (q)}$
and with equations of motion given in Proposition 5.6.

\proclaim
{Corollary 6.5}  Let $(q_0,p_0,s_0) \in TU \times \fg_{red}$ and
suppose $s_0 =  Ad_{g(\xi_{0})^{-1}} \xi_{0}$ where 
$\xi_{0}\in {\Cal U} \cap \fh^{\perp}.$  Then the Hamiltonian
flow generated by ${\Cal H}_{0}$ with initial condition
$(q(0),p(0),s(0)) = (q_0,p_0,s_0)$ is given by
$$\eqalign{
       & q(t) =\, log\,d(t), \cr
       & s(t) =\, Ad_{\bigl({\widetilde k}_{+}(t)\,
                  g \bigl(Ad_{{\widetilde k}_{+}(t)^{-1}} s_{o}\bigr)
       \bigr)^{-1}} s_{0},\cr
       & p(t) =\, Ad_{\bigl({\widetilde k}_{+}(t)\,
                  g\bigl(Ad_{{\widetilde k}_{+}(t)^{-1}} s_{o}\bigr)
         \bigr)^{-1}} (p_0 -R^{-}(q_0)s_0)+ R^{-}(q(t)) s(t).\cr}
\eqno(6.32)
$$
where $ {\widetilde k}_{+}(t) = g(\xi_{0})^{-1} k_{+}(t) g(\xi_{0})$
and $k_{+}(t)$, $d(t)$ are as in Theorem 6.4.
\endproclaim

\demo
{Proof} We shall obtain the Hamiltonian flow generated by ${\Cal H}_{0}$ 
by reduction.  Using the relation 
$\phi^{red}_t \circ \pi_{0} = \pi_{0} \circ \phi_{t} \circ i_{0}$
from Theorem 3.2 (c), we have
$\phi^{red}_t (q_0,p_0,s_0) = (q(t),p(t), Ad_{g(\xi(t))^{-1}} \xi(t))$
where $q(t)$, $p(t)$ $\xi(t)$ are given by the expressions in 
Theorem 6.4.  Thus
$$
\aligned
       s(t) = & Ad_{g(\xi(t))^{-1}} \xi(t) \\
            = & Ad_{g\bigl(Ad_{k_{+}(t)^{-1}}\xi_{0}\bigr)^{-1}}
                Ad_{k_{+}(t)^{-1} g(\xi_{0})} s_{0}\\
            = & Ad_{\bigl({\widetilde k}_{+}(t)
                  g\bigl(Ad_{{\widetilde k}_{+}(t)^{-1}} s_{o}\bigr)\bigr)
               ^{-1}} s_{0}
\endaligned
$$
where we have used the $H$-equivariance of the map $g$.
To express $p(t)$ in the desired form, introduce the gauge
transformation of $L$: ${\widetilde L} (q,p,s) = Ad_{g(\xi)^{-1}} L(q,p,\xi)
= p - R^{-}(q) s$, where as before, $s=Ad_{g(\xi)^{-1}} \xi$.
Then 
$$
\aligned
   {\widetilde L} (q(t),p(t),s(t)) = & Ad_{g(\xi(t))^{-1}}Ad_{k_{+}(t)^{-1}}
      L(q_0,p_0,\xi_0) \\
 = & Ad_{\bigl({\widetilde k}_{+}(t) g\bigl(Ad_{{\widetilde k}_{+}(t)^{-1}} 
     s_{o}\bigr)\bigr)^{-1}}
      (p_0 - R^{-}(q_0)s_0).
\endaligned
$$
But ${\widetilde L} (q(t),p(t),s(t))$ is also equal to
$p(t) - R^{-}(q(t))s(t)$.  By equating the two expressions,
we obtain the desired expression for $p(t)$, as claimed.      
\pf
\enddemo

\noindent {\bf Remark 6.6} (a)It is easy to show that the element
${\widetilde k}_{+}(t) = g(\xi_{0})^{-1} k_{+}(t) g(\xi_{0})$
depends only on $s_0$, and not on the particular element
$\xi_{0} \in {\Cal U} \cap \fh^{\perp}$ for which
$Ad_{g(\xi_{0})^{-1}} \xi_{0} = s_0$.  Indeed, from the
factorization $e^{L(q_0,p_0,\xi_0)} = k_{+}(t) k_{-}(t)^{-1}$,
we see that if we replace$\xi_0$ by $Ad_{h}\xi_{0}$, $h\in H$, 
then the factors $k_{\pm}(t)$ will be replaced by
$h k_{\pm}(t) h^{-1}$.  As $g(Ad_{h}\xi_0) = h g(\xi_0)$, our
assertion easily follows.  Note that this is exactly the reason why we
have chosen to express $s(t)$ and $p(t)$ in the form given in
the above Corollary.
\smallskip
\noindent (b) In \c{L1}, we introduced a family of hyperbolic spin
Ruijsenaars-Schneider models on the coboundary dynamical Poisson
groupoids $(\Gamma = U\times G\times U, \{\,\cdot, \cdot \,\}_{R})$
associated to the same $R$'s  which we use here. Recall that these
are generated by nonzero multiples of 
$H_i = Pr_2^{*}\chi_i$, $i = 1,\cdots, N,$  where $Pr_2$ in this
present case denotes projection onto the second factor of $\Gamma$,
and $\chi_1,\ldots, \chi_N$ are the characters of the irreducible
representations corresponding to the fundamental weights 
$\omega_1,\ldots, \omega_N$ \c{St}.  If we take the Hamiltonian
$H_i$, say, then its Hamiltonian flow on the gauge group bundle
${\Cal I}\Gamma$ is defined by the equation
$$
\aligned
       & {d\over dt} (q,g,q) \\
   = \,& \bigl(-{1\over 2} \Pi_{\fh}\,D\chi_{i} (g),
         {1\over 2} T_e r_g\,R(q)(D\chi_{i}(g)) - {1\over 2}
         T_e l_g\,R(q)(D\chi_{i}(g)),-{1\over 2} \Pi_{\fh}\,D\chi_{i} (g)
         \bigr)
\endaligned
$$
where $D\chi_{i}(g)$ is the right gradient of $\chi_i$.  In this
case, the factorization problem on $\Gamma$ (from \c{L1} and our
analysis above) gives
$$
\aligned
       & e^{-t D\chi_{i} (g_0)} \\
        =\, & \bn^{-}(k_{+}(t))(\bl^{-}(k_{+}(t))
           e^{q(t)}\bl^{-}(k_{+}(t))^{-1}e^{-q_{0}})
           \bn^{+}(k_{-}(t))^{-1} 
\endaligned
$$
using the same notation as before (of course, the $k_{\pm}$ here
are different from the ones above).  If for small $t$, 
$n_{\pm}(t)\in N^{\pm}_{\pi^{\prime}}$, $g(t)\in G_{\pi^{\prime}}$
are the unique solution of the factorization problem
$$e^{-t D\chi_{i} (g_0)} = n_{-}(t) g(t) n_{+}(t)^{-1}$$
satisfying $n_{\pm}(0) = g(0) =1$, then
$$\bn^{-} (k_{+}(t)) = n_{-}(t), \quad \bn^{+}(k_{-}(t)) = n_{+}(t).$$
Consequently, the factorization problem reduces to 
$$g(t) e^{q_{0}} = \bl^{-}(k_{+}(t))
           e^{q(t)}\bl^{-}(k_{+}(t))^{-1}$$
and therefore the solution proceeds as before.  Finally, we can 
write down the Hamiltonian flow on ${\Cal I}\Gamma$ using the
formula from \c{L1}, namely,
$$
\aligned
       &(q(t),g(t),q(t))\\
=\,&(q_0, k_{\pm} (t), q(t))^{-1}(q_0,g_0,q_0)(q_0, k_{\pm} (t), q(t)).
\endaligned
$$

\subsubhead (b) The spin Toda lattices
\endsubsubhead

\bigskip

In this final subsection, we shall discuss the solution of the
spin Toda lattices.  In this case, we have 
$$Im {\Cal R}^{\pm} = \bigcup_{q\in \fh} \lbrace 0_{q} \rbrace \times
{\frak b}^{\mp}\times \fh  \eqno(6.33)$$
where ${\frak b}^{\pm}$ are the opposing Borel subalgebra of
$\fg$ introduced at the beginning of the section.  Let $B^{\pm}$ be
the simply-connected Lie subgroups of $G$ which integrate ${\frak b}^{\pm}$,
then the unique source-simply connected Lie groupoid of 
$Im {\Cal R}^{\pm}$ are given by
$$ \Gamma_{\pm} = \fh\times B^{\mp}\times \fh.\eqno(6.34)$$
Now, the ideals ${\Cal I}^{\pm}$ of (4.7) in the present case are:
$${\Cal I}^{\pm} = \bigcup_{q\in \fh} \lbrace 0_q \rbrace
\times {\frak n}^{\mp} \times \lbrace 0 \rbrace\eqno(6.35)$$
where ${\frak n}^{\pm} = \sum_{\alpha \in \Delta ^{\pm}} \fg_{\alpha}$.
We shall denote by $N^{\pm}$ the simply-connected Lie subgroups of
$G$ with $Lie (N^{\pm}) = {\frak n}^{\pm}$.  To cut the story short,
we have the following result when we go through the analysis.

\proclaim
{Theorem 6.7} Let $(x_0,p_0, \eta_0)\in T\fh \times \fg \simeq S^{*}$
and let $n_{\pm}(t) \in N^{\pm}$, $h(t)\in H$ be the unique
solution of the factorization problem
$$e^{t{\bold L}(x_0,p_0,\eta_0)} = n_{-}(t) h(t) n_{+}(t)^{-1}\eqno(6.36)$$
(valid for $0 \le t < T$ for some $T > 0$) satisfying $n_{\pm}(0) =
h(0) = 1$. Then the Hamiltonian flow on $ T\fh \times \fg \simeq S^{*}$
generated by the Hamiltonian
$$
\aligned
{\Cal H}^{s} (x,p,\eta) =& {1 \over 2} \sum_{i} p_{i}^{2} + 
          {1 \over 8} \sum_{i} \eta_{i}^{2} + {1 \over 2} \sum_{i}
           p_{i}\eta_{i} \\
           & -\sum_{\alpha \in \pi^{\prime}} \eta_{\alpha}
           \eta_{-\alpha} e^ {-\alpha(x)}
\endaligned
$$
with initial condition $(x(0), p(0),\eta(0)) = (x_0,p_0,\eta_0)$ is 
given by
$$\eqalign{& x(t) =\, x_0 + log\,h(t), \cr
           & \eta(t) =\, Ad_{e^{-{1\over 2} log\,h(t)}} \eta_{0}, \cr
           & p(t) = Ad_{k_{\pm}(t)^{-1}} {\bold L} (x_0,p_0,\eta_0)
              -{1\over 2} \Pi_{\fh}\,\eta_0
             -\sum_{\alpha\in \pi}
              e^{-{1\over 2} \alpha (log\,h(t))} (\eta_{0})_{-\alpha}
             e_{\alpha}\cr
           &\,\,\,\,\,\,\,\,\,\, + \sum_{\alpha\in \pi^{\prime}}
             e^{-\alpha(x_0 +{1\over 2} log\,h(t))}
             (\eta_0)_{-\alpha} e_{\alpha}\cr}\eqno(6.37)$$
where 
$$k_{\pm}(t)=\, n_{\mp}(t) e^{\pm {1\over 2} log\,h(t)}.\eqno(6.38)$$
\endproclaim

\demo
{Proof}  The expression for $x(t)$ is clear if we write down the
expression analogous to (6.21) and compare that with the factorization
in (6.36).
On the other hand, it  is clear from the same expression that 
$k_{+}(t) = n_{-}(t) b_{-}(t)$ where
$b_{-}(t) \in H$ is to be determined from the condition given
in (4.19).  If we spell this out, we find that $b_{-}(t)$ satisfies the
equation
$$\dot b_{-}(t) = {1\over 2}\, T_{e}\, l_{b_{-}(t)}\, \dot x(t)$$
with $b_{-}(0) = 1$.  Solving the equation explicitly, we have
 $b_{-}(t) = e^{{1\over 2} (x(t) - x_0)}$.  Now, in order to solve
the equation for $\eta$ in (5.24), the crucial point to note is that
we can rewrite this equation  as $\dot \eta = [\,\eta, {1\over 2} \Pi_{\fh}
{\bold L} (x,p,\eta)]$.
Finally, we can obtain the formula for $p(t)$ by equating the 
following two expressions for ${\bold L} (x(t),p(t),\eta(t))$, namely,
${\bold L} (x(t),p(t),\eta(t))= Ad_{k_{\pm}(t)^{-1}} {\bold L} 
(x_0,p_0,\eta_0)$ and
$$\eqalign{\bL(x(t),p(t),\eta(t))=\, & p(t) + {1\over 2} \Pi_{\fh}\,\eta_0 + 
     \sum_{\alpha \in \pi} \eta_{\alpha}(t)
      e_{\alpha} -\sum_{\alpha \in \pi^{\prime}} e^{-\alpha(x(t))}
      \eta_{-\alpha}(t) e_{-\alpha}.\cr}$$
This completes the proof.
\pf
\enddemo

The solution of the family of Toda lattices in Corollary 5.12 is 
now straightforward.  We shall leave the easy details to the reader.

\newpage

\head
{\bf Appendix}
\endhead

\bigskip
\subsubhead  Proof of Lemma 4.1
\endsubsubhead
\bigskip

From the definition of ${\Cal R}$ and the expression for 
$[\,\cdot,\cdot\,]_{A\Gamma}$, $[\,\cdot,\cdot,\,]_{A^{*}\Gamma}$, we
have
$$
\aligned
       & [{\Cal R}(0,A,Z),{\Cal R}(0,A',Z')]_{A\Gamma} (q)-
        {\Cal R}[(0,A,Z),(0,A',Z')]_{A^*\Gamma} (q)\\
    =\, & (0_q,{\Cal A}, {\Cal Z})
\endaligned
$$
where (after the obvious cancellations)
$$
\aligned
      {\Cal A} =\, & [R(q)A(q), R(q)A'(q)] + < dR(q)(\cdot)A(q),A'(q)>
       +\Bigl\{R(q)ad^*_{R(q)A(q)} A'(q) \\
      &+ dR(q)\iota^*A(q) (A'(q))- dR(q) ad^{*}_{Z(q)} q (A'(q)) -
         [Z(q), R(q)A'(q)]\\
      &- R(q) ad^{*}_{Z(q)}A'(q) -(A\leftrightarrow A', Z\leftrightarrow
        Z')\Bigr\}
\endaligned
$$
and
$${\Cal Z} =  \iota^{*} ad^{*}_{R(q)A(q)} A'(q) 
       -\iota^{*} ad^{*}_{R(q)A'(q)} A(q)
       + ad^{*}_{<dR(q)(\cdot)A(q),A'(q)>} q.$$
Since $R$ is $H$-equivariant, we can show that ${\Cal Z} =0$ and
$$dR(q) ad^{*}_{Z(q)} q (A'(q)) +[Z(q), R(q)A'(q)] +R(q) ad^{*}_{Z(q)}A'(q)
    =\, 0.$$
Therefore, the expression for ${\Cal A}$ becomes
$$
\aligned
    {\Cal A} =\, &[R(q)A(q), R(q)A'(q)]+< dR(q)(\cdot)A(q),A'(q)>\\
               &+ R(q)(ad^*_{R(q)A(q)}a'(q)-ad^*_{R(q)A'(q)}A(q))\\
               &+ dR(q)\iota^*A(q)(A'(q)) - dR(q)\iota^*A'(q)(A(q))\\
             =\, & - [K(A(q)), K(A'(q))],
\endaligned
$$
as desired.

\enddocument